\documentclass{article}

\usepackage{arxiv}
\usepackage{url}
\usepackage{fancyhdr}
\usepackage{bbm}
\usepackage{tabularx}
\newcolumntype{Y}{>{\centering\arraybackslash}X}
\newcolumntype{Z}{>{\raggedright\arraybackslash}X}
\usepackage{longtable}
\usepackage{multirow}
\usepackage{diagbox}
\usepackage{algorithm, algorithmic}
\usepackage{subfigure}
\usepackage{makecell}
\usepackage[T1]{fontenc}

\RequirePackage{algorithm}
\usepackage[backref]{hyperref}

\usepackage{cite}


\usepackage{xcolor}
\definecolor{lightgray}{gray}{0.893}
\usepackage{colortbl}

\usepackage{mathtools} 
\usepackage{booktabs} 
\usepackage{tikz} 
\usepackage{bm}
\usepackage{amssymb,amsmath,amsthm}

\newtheorem{definition}{Definition}
\newtheorem{theorem}{Theorem}

\title{A Survey of Progress on Cooperative Multi-agent Reinforcement Learning
in Open Environment}

\author{%
 Lei Yuan, Ziqian Zhang, Lihe Li, Cong Guan, {Yang Yu}\thanks{Corresponding Author}\\
  National Key Laboratory for Novel Software Technology, Nanjing University\\
  School of Artificial Intelligence, Nanjing University\\
  \texttt{\{yuanl, zhangzq, lilh, guanc\}@lamda.nju.edu.cn}, \texttt{yuy@nju.edu.cn}
}

\date{}
  
\begin{document}
\pagestyle{mystyle}
\maketitle

\begin{abstract}
Multi-agent Reinforcement Learning (MARL) has gained wide attention in recent years and has made progress in various fields. Specifically, cooperative MARL focuses on training a team of agents to cooperatively achieve tasks that are difficult for a single agent to handle. It has shown great potential in applications such as path planning, autonomous driving, active voltage control, and dynamic algorithm configuration. One of the research focuses in the field of cooperative MARL is how to improve the coordination efficiency of the system, while research work has mainly been conducted in simple, static, and closed environment settings. 
To promote the application of artificial intelligence in real-world, some research has begun to explore multi-agent coordination in open environments. These works have made progress in exploring and researching the environments where important factors might change.
However, the mainstream work still lacks a comprehensive review of the research direction. In this paper, starting from the concept of reinforcement learning, we subsequently introduce multi-agent systems (MAS), cooperative MARL, typical methods, and test environments. Then, we summarize the research work of cooperative MARL from closed to open environments, extract multiple research directions, and introduce typical works. Finally, we summarize the strengths and weaknesses of the current research, and look forward to the future development direction and research problems in cooperative MARL in open environments.
\end{abstract}

\section{Introduction}
As a sub-branch of machine learning, reinforcement learning (RL) \cite{sutton2018reinforcement} is an effective method for solving sequential decision-making problems. Compared to supervised learning and unsupervised learning, RL learns from interactions. In the paradigm of RL, an agent interacts with the environment and continuously optimizes its policy based on the rewards or penalties it receives from the environment. Due to its similarity to the way humans acquire knowledge, RL is considered as one of the approaches to achieving Artificial General Intelligence (AGI) \cite{goertzel2007artificial}. Early work in RL relied on handcrafted features input into linear models for value estimation and approximation, which performed poorly in complex scenarios.
In the past decade, deep RL has achieved remarkable achievements in various fields with the flourishing development of deep learning \cite{lecun2015deep}. For example, Deep Q-Network (DQN) \cite{mnih2013playing} surpassed professional human players in Atari video games. AlphaGo \cite{silver2016mastering} defeated the world champion Go player, Lee Sedol. AlphaStar \cite{vinyals2019grandmaster} defeated top human professional players in the imperfect information real-time policy game StarCraft II. OpenAI Five \cite{berner2019dota} performed well in the multiplayer real-time online game Dota 2. AI-Suphx \cite{li2020suphx} also achieved significant results in multiplayer imperfect information Mahjong games.
In addition, the application scope of RL has gradually expanded from the games to various domains in real life, including industrial manufacturing, robotic control, logistics management, defense and military affairs, intelligent transportation, and intelligent healthcare, greatly promoting the development of artificial intelligence \cite{li2017deep, shakya2023reinforcement}. For example, ChatGPT \cite{liu2023summary}, which has recently received widespread attention, also uses RL techniques for optimization. In recent years, under the trend of applying artificial intelligence to scientific research (AI4Science) \cite{wang2023scientific}, RL has also shined in many fundamental scientific fields. For example, DeepMind achieved nuclear fusion control \cite{DBLP:journals/nature/DegraveFBNTCEHA22} with the application of RL. AlphaTensor has also applied RL into discovering matrix multiplication \cite{fawzi2022discovering}.

At the same time, many real-world problems are shown to be large-scale, complex, real-time, and uncertain. Formulating such problems as the single-agent system is inefficient and inconsistent with real conditions, while modeling them as multi-agent systems (MAS)~\cite{dorri2018multi} problems is often more suitable. Furthermore, multi-agent coordination has been applied into dealing with many complex problems, such as autonomous driving cars, intelligent warehousing systems, and sensor networks.
Multi-agent reinforcement learning (MARL) \cite{yang2020overview, oroojlooy2023review, marl-book} provides strong support for modeling and solving these problems. In MARL, a team of agents learns a joint cooperative policy to solve the tasks through interactions with the environment. Compared to traditional methods, the advantages of MARL lie in its ability to deal with environmental uncertainty and learn to solve unknown tasks without requiring excessive domain knowledge. In recent years, the combination of deep learning and MARL has produced fruitful results \cite{gronauer2022multi}, and many algorithms have been proposed and applied to solve complex tasks.
However, MARL also brings new challenges. On the one hand, the environment where the MAS exists is often {partially observable}, and the individual cannot obtain global information from its local observations. This means that independently learning agents struggle to make optimal decisions \cite{zhu2022survey}. On the other hand, since other agents are also learning simultaneously, the policies will change accordingly. From the perspective of an individual agent, the environment is {non-stationary}, and convergence cannot be guaranteed \cite{papoudakis2019dealing}. In addition, cooperative MAS often receive shared rewards only, and how to allocate these rewards to provide accurate feedback for each agent (a.k.a, {credit assignment}), thereby enabling efficient learning of cooperation and ultimately maximizing system performance, is one of the key challenges \cite{wang2021towards}. Finally, as the number of agents in a MAS increases, the search space faced in solving RL problems will exponentially expand, making policy learning and search extremely difficult, bring the {scalability} issue. Therefore, organizing efficient policy learning is also a major challenge at present \cite{zhang2011scaling, christianos2021scaling}.
\begin{figure*}
\centering
\includegraphics[width=1\linewidth]{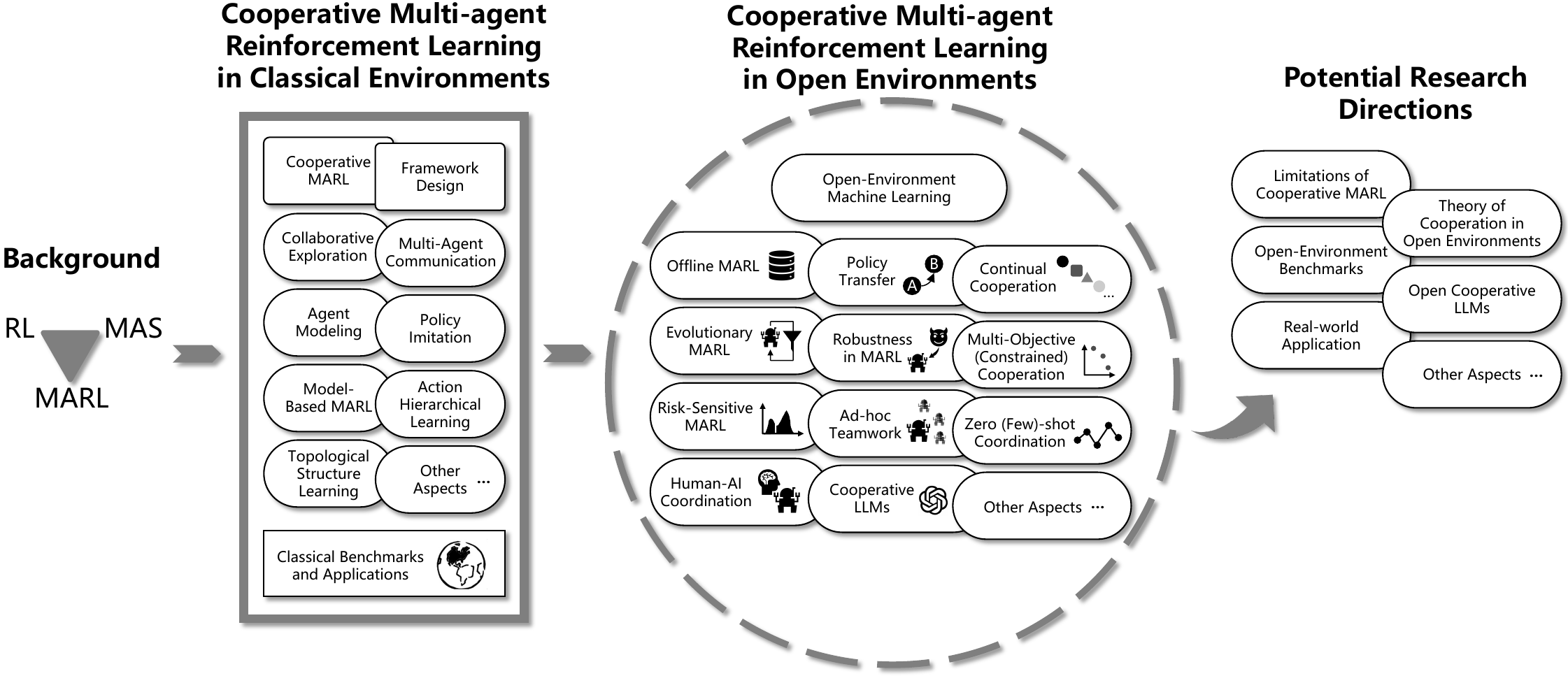}
\caption{Framework of the survey}\label{frameworksurvey}
\end{figure*}

To address the aforementioned challenges, a large amount of work is currently being conducted from multiple aspects, and surprising achievements have been made in many task scenarios~\cite{marl-book}. Cooperative MARL has demonstrated superior performance compared to traditional methods in tasks such as path planning \cite{sartoretti2019primal}, active voltage control \cite{wang2021multi}, and dynamic algorithm configuration \cite{xue2022multi}. Researchers have designed many algorithms to promote cooperation among agents, including policy gradient based methods such as MADDPG \cite{maddpg} and MAPPO \cite{mappo}, value based methods such as VDN \cite{vdn} and QMIX \cite{qmix}, and other methods that leverage the powerful expressive power of Transformers to enhance coordination capabilities, such as MAT \cite{wen2022multiagent}. These methods have demonstrated excellent cooperation ability in many tasks, such as SMAC \cite{pymarl}, Hanabi, and GRF \cite{mappo}. In addition to the above methods and their respective variants, researchers have also conducted in-depth exploration and research on cooperative MARL from other perspectives, including alleviating partial observability under distributed policy execution settings through efficient communication \cite{zhu2022survey}, offline deployment of policies \cite{zhang2023discovering}, world model learning in MARL \cite{wang2022model}, and research on training paradigms \cite{lyu2021contrasting}.

Traditional machine learning research is typically conducted under the assumption of classical closed environments, where crucial factors in the learning process remain constant. Nowadays, an increasing number of tasks, especially those involving open environment scenarios, may experience changes in essential learning factors. Clearly, transitioning from classical to open environments poses a significant challenge for machine learning.
For data-driven learning tasks, data in open environments accumulates online over time, such as in the form of data streams, making model learning more challenging.
Machine learning in open environments~\cite{parmar2023open, zhou2022open} has gained application prospects in many scenarios, gradually attracting widespread attention. Current research in open environment machine learning includes category changes, feature evolution, data distribution changes, and variations in learning objectives. Correspondingly, some works in the field of RL have started focusing on tasks in open environments. Key areas of research include trustworthy RL~\cite{Xu2022TrustworthyRL}, environment generation and policy learning~\cite{wang2020enhanced}, continual RL~\cite{khetarpal2022towards}, RL generalization capabilities~\cite{kirk2023survey}, meta-RL~\cite{beck2023survey}, and sim-to-real policy transfer~\cite{zhao2020sim}.

Compared to single-agent reinforcement learning (SARL), multi-agent scenarios are more complex and challenging. Currently, there is limited research on cooperative MASs in open environments, with some efforts focusing on robustness in multi-agent settings~\cite{guo2022towards}. These works describe problems and propose algorithmic designs from different perspectives~\cite{van2020robust, zhang2020robustmarl,hu2021robust, xue2022mis}. Additionally, addressing the challenges of open-team MARL, some works introduce settings such as Ad-Hoc Teamwork (AHT), Zero-Shot Coordination (ZSC), and Few-Shot Teamwork (FST) to tackle it~\cite{mirsky2022survey,treutlein2021new,fosong2022few}. Although these works have achieved success in some task scenarios, they still fail to align well with most real-world applications, leaving room for substantial improvement in practical effectiveness.
Regarding MARL, there exist some review works, such as those on multi-agent systems~\cite{dorri2018multi}, MARL~\cite{shoham2003multi,shoham2007if,busoniu2008comprehensive,hernandez2019survey, nguyen2020deep, yang2020overview,zhang2021multi}, agent modeling in multi-agent scenarios~\cite{albrecht2018autonomous}, non-stationarity handling in multi-agent settings~\cite{papoudakis2019dealing}, multi-agent transfer learning~\cite{da2019survey}, cooperative MARL~\cite{doran1997cooperation,panait2005cooperative,oroojlooy2023review}, model-based multi-agent learning~\cite{wang2022model}, causal MARL~\cite{grimbly2021causal}, and multi-agent communication~\cite{zhu2022survey}. Additionally, some works provide comprehensive analyses of open machine learning~\cite{parmar2023open,zhou2022open,kim2023open}.
Although the mentioned works provide reviews and summaries in various aspects of MARL or open environment machine learning, there is currently no systematic review specifically focusing on cooperative MARL in open environments. Considering the potential and value of cooperative MARL in solving complex coordination problems in real environments, this paper aims to describe recent advances in this field. The subsequent arrangement of this paper is shown in Figure~\ref{frameworksurvey}. We first introduce the background relevant to this paper, including basic knowledge of RL, common knowledge and background of MASs to MARL. Next, we introduce cooperative MARL in classical closed environments, covering specific definitions, current mainstream research content, and common testing environments and application cases. Following that, we introduce cooperative MARL in open environments, specifically including common research directions and content from closed machine learning and reinforcement learning to cooperative multi-agent scenarios. Finally, we summarize the main content of this paper, provide a prospect on cooperative MARL in open environments, and aim to inspire further research and exploration in this direction.
\section{Background}
\subsection{Reinforcement Learning}

Reinforcement learning~\cite{sutton2018reinforcement} aims to guide an agent to learn the appropriate actions based on the current state, i.e., learning a mapping from observed state to action, in order to maximize the cumulative numerical reward feedback from the environment. The environment provides reward information based on the current state and the action taken by the agent. The agent does not know the optimal actions in advance and must discover actions that yield the highest cumulative reward through trial and error. In a standard RL scenario, the agent interacts with the environment by observing states and taking actions. At each time step, the agent receives an observation of the current state, selects an action based on that observation, and the execution will change the state of the environment and lead to a reward signal. The goal of the agent is to execute a sequence of actions that maximizes the cumulative numerical reward.
\subsubsection{Formulation of the problem}
\begin{figure*}
\centering
\includegraphics[width=0.6\linewidth]{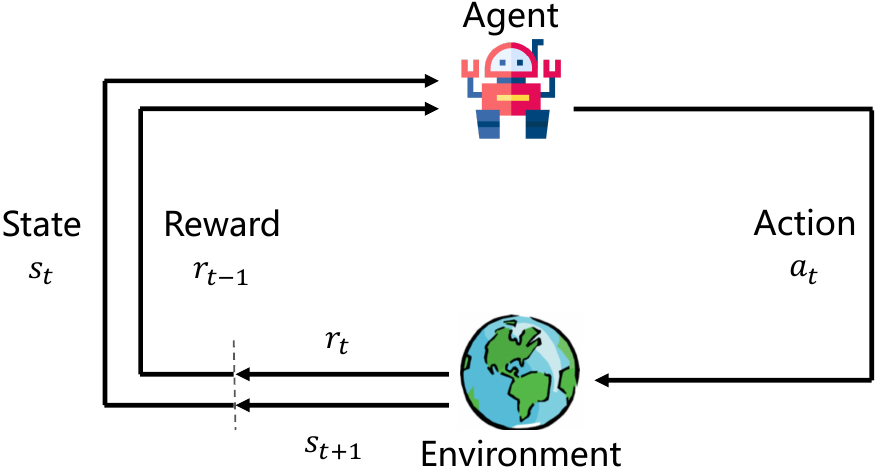}
\caption{Illustration of reinforcement learning.}\label{mdp}
\end{figure*}

RL~\cite{sutton2018reinforcement} is a sub-branch of machine learning that differs from classical supervised and unsupervised learning in that it learns through interaction. In RL, an agent interacts with an environment, constantly optimizing its policy based on the rewards or punishments received from the environment. RL consists of four main components: agent, state, action, and reward (Figure \ref{mdp}). The goal of RL is to maximize cumulative rewards. The agent must explore and learn through trial and error to find the optimal policy. This process can be modeled as a Markov Decision Process (MDP).

\begin{definition}[Markov Decision Process]\label{def_MarkovDecisionProcess}
A Markov Decision Process is defined by either the five-tuple $\langle \mathcal{S},\mathcal{A},P,R,\gamma\rangle$ (infinite process) or $\langle \mathcal{S},\mathcal{A},P,R,T\rangle$ (finite process), where:
\begin{itemize}
\item $\mathcal{S}$ is the set of states,
\item $\mathcal{A}$ is the set of actions,
\item $P: \mathcal S\times \mathcal A \times \mathcal S\to [0,1]$ is the state transition probability function: $P (s'\mid s, a)=\Pr [S_{t+1}=s'\mid S_t=s,A_t=a]$,
\item $R:\mathcal S\times \mathcal A \to
\mathbb R$ is the reward function: $R(s, a)=\mathbb{E}[r_{t}\mid S_t=s,A_t=a]$,
\item $\gamma\in[0, 1]$ is the discount factor, $T$ is the maximum horizon.
\end{itemize}
\end{definition}

The agent is free to choose deterministic action to perform in a given state or to sample actions from some certain probability distributions. The emphasis here is on "free choice," which refers to the agent's autonomy in changing action selection. The fixed selection of actions is referred to as a "policy".
Specifically, at any time step $t$, the agent observes the current state $s_t$ from the environment and then performs action $a_t$. This action causes the environment to transition to a new state $s_{t+1}$ based on the transition function $s_{t+1}\sim P(\cdot|s_t,a_t)$ and the agent receives a reward signal $r_t=R(s_t,a_t)$ from the environment. Without loss of generality, we mainly discusses infinite MDP in this paper, where the goal of the agent is to find the optimal policy to maximize cumulative rewards. Mathematically, the above process can be summarized as finding a Markov, time-independent, and stationary policy function $\pi:\mathcal{S}\rightarrow \Delta(\mathcal{A})$ that guides the agent to execute appropriate sequential decisions to maximize cumulative rewards, with the optimization objective defined as follows: 
\begin{equation}
\begin{aligned}
\mathbb{E}^\pi\left[\sum_{t = 0}^\infty \gamma^t R(s_t,a_t)\mid s_0 \right],
\end{aligned}
\label{intro:1}
\end{equation}
where the operator $\mathbb{E}^{\pi}[\cdot]$ computes the expectation over the distribution of sequences $\tau=(s_0,a_0,s_1,a_1,\ldots)$ generated by the policy $\pi(a_t|s_t)$ and the state transition $P(s_{t+1}|s_t, a_t)$.

Based on the objective in Equation \ref{intro:1}, the state-action value function, denoted as the $Q$-function, can be defined under the guidance of policy $\pi$. The $Q$-function represents the expected cumulative reward after taking action $a$ based on state $s$ under policy $\pi$. Additionally, the state value function $V$ describes the expected cumulative reward based on state $s$: \begin{equation}
\begin{aligned}
Q^{\pi}(s,a)&=\mathbb{E}^{\pi}\left[\sum_{t= 0}^\infty\gamma^t R(s_t,a_t)\mid s_0=s,a_0=a \right], \\
V^{\pi}(s)&=\mathbb{E}^{\pi}\left[\sum_{t= 0}^\infty\gamma^t R(s_t,a_t)\mid s_0=s \right].
\end{aligned}
\label{sec1:v}
\end{equation} 
Clearly, the relationship between the state-action function $Q$ and the state value function $V$ can be expressed as $V^{\pi}(s)=\mathbb{E}_{a\sim \pi(\cdot|s)}\left[Q^\pi(s,a) \right]$ and $Q^\pi(s,a)=\mathbb{E}_{s'\sim P(\cdot|s,a)}\left[R(s,a)+V^\pi(s') \right]$. Based on the definitions of these two types of value functions, the learning goal of RL in Markov decision processes can be represented as finding the optimal policy $\pi_*$ that maximizes the value function: $\pi_*=\arg\max_{\pi} V^{\pi}(s), \forall s$.

\subsubsection{Value based Reinforcement Learning}

For the MDP with finite state and action space, there exists at least one deterministic stationary optimal policy that maximizes the cumulative reward. Value based RL methods require the construction and estimation of a value function, which is subsequently used to guide action selection and execution, deriving the corresponding policy. Most of these algorithms do not directly use the $V$-function but require the use of the $Q$-value function to maximize the $Q$-value corresponding to Equation~\ref{sec1:v}. The optimal policy induced by the $Q$-function through greedy action selection can be expressed as $\pi^*(a|s)=\mathbbm{1}\{a=\arg\max_aQ^*(s,a)\}$, where $\mathbbm{1}\{\cdot\}$ is the indicator function. The classical $Q$-learning algorithm, through temporal-difference (TD) learning, updates the agent's function $\hat{Q}$-value to approximate the optimal function $Q^*$~\cite{watkins1992q}, with the update process defined as follows:
\begin{equation}
\begin{aligned}
\hat{Q}&(s_t,a_t) \leftarrow \hat{Q}(s_t,a_t)+
\alpha \cdot \overbrace{\left(\underbrace{r_t+\gamma\max_{a_{t+1}\in \mathcal{A}} \hat{Q}(s_{t+1},a_{t+1})}_{\mbox{temporal-difference target}}-\hat{Q}(s_t,a_t) \right)}^{\mbox{temporal-difference error}}.
\end{aligned}
\end{equation}

In theory, given the Bellman optimality operator $\mathcal{B}^*$, the update of $Q$-value function can be defined as~\cite{Tesauro95}:
\begin{equation}
\begin{aligned}
(\mathcal{B}^*Q)&(s,a)= \sum_{s'}P(s'|s,a)\left[R(s,a)+\gamma\max_{a'\in \mathcal{A}}Q(s',a') \right].
\end{aligned}
\label{sec1:bellman-q}
\end{equation}
To establish the optimal $Q$-value function, the $Q$-learning algorithm uses the fixed-point iteration of the Bellman equation to solve the unique solution for $Q^*(s,a)=(\mathcal{B}^* Q^*)(s,a)$. In practice, when the world model~\cite{luo2022survey} is unknown, state-action pairs are discretely represented, and all actions can be repetitively sampled in all states, the mentioned $Q$-learning method guarantees convergence to the optimal solution.

However, real-world problems may involve continuous, high-dimensional state-action spaces, making the above assumptions usually inapplicable. In such cases, agents need to learn a parameterized $Q$-value function $Q(s,a|\theta)$, where $\theta$ is the parameter instantiation of the function. To update the $Q$-function, the agent needs to collect samples generated during interaction with the environment in the form of tuples $(s,a,r,s')$. Here, the reward $r$ and the state $s'$ at the next time step follow feedback from the environment based on the state-action pair $(s,a)$. At each iteration, the current $Q$-value function $Q(s,a|\theta)$ can be updated as follows~\cite{bradtke1996linear}:
\begin{equation}
    \begin{aligned}
        y^Q&=r+\gamma \max_{a'\in \mathcal{A}} Q(s',a'|\theta),\\ 
	\theta&\leftarrow\theta+\alpha \left(y^Q-Q(s,a|\theta)\right)\nabla_{\theta}Q(s,a|\theta) ,
    \end{aligned}
    \label{sec1:q-loss}
\end{equation}
where $\alpha$ is the update rate.

The $Q$-learning approach in Equation~\ref{sec1:q-loss} can directly use a neural network $Q(s,a|\theta)$ to fit and converge to the optimal $Q^*$-value, where the parameters $\theta$ are updated through stochastic gradient descent (or other optimization methods). However, due to the limited generalization and reasoning capabilities of neural networks, $Q$-networks may exhibit unpredictable variations in different locations of the state-action space. Consequently, the contraction mapping property of the Bellman operator in Equation~\ref{sec1:bellman-q} is insufficient to guarantee convergence. Extensive experiments show that these errors propagate with the propagation of online update rules, leading to slow or even unstable convergence. Another disadvantage of using function approximation for $Q$-values is that, due to the action of the $\max$ operator, $Q$-values are often overestimated. Therefore, due to the risks of instability and overestimation, special attention must be paid to ensuring an appropriate learning rate. Deep Q Network (DQN) algorithm proposed by literature~\cite{MnihKSRVBGRFOPB15} addresses these challenges by incorporating two crucial factors: a target $Q$-network and experience replay buffer. Specifically, DQN instantiates a target $Q$-network with parameters $\theta^-$ to compute the temporal-difference target. Simultaneously, it uses an experience replay buffer $\mathcal{D}$ to store sampled sequences, ensuring both sample utilization and, during training, mitigating the correlation of sampled data through independent and identically distributed sampling. The optimization objective of DQN can be expressed as:
\begin{equation}
    \begin{aligned}    
	\min_{\theta} \mathbb{E}_{(s,a,r,s')\sim \mathcal{D}}[(r+\gamma    \max_{a'\in\mathcal{A}}Q(s',a'|\theta^-)
 -Q(s,a|\theta) )^2 ],
    \end{aligned}
 \label{sec1:dqn_loss}
\end{equation}
The target $Q$-network $Q(s,a|\theta^-)$ is periodically updated to synchronize with the parameters of the $Q$-network. Through these methods and two crucial factors, stable training of the $Q$-network can be achieved.

Furthermore, many recent works have made additional improvements to DQN. In \cite{WangSHHLF16}, the value function and advantage function (defined as $A(s,a)=Q(s,a)-V(s)$) are decoupled using a neural network structure, enhancing learning performance. In \cite{HasseltGS16}, a specific update scheme (a variant of Equation~\ref{sec1:dqn_loss}) reduces the overestimation of $Q$-values while improving learning stability. Parallel learning~\cite{BellemareDM17} or the use of unsupervised auxiliary tasks~\cite{JaderbergMCSLSK17} contributes to faster and more robust learning. In \cite{PritzelUSBVHWB17}, a differentiable memory module can integrate recent experiences by interpolating between Monte Carlo value estimates and off-policy estimates.

\subsubsection{Policy Gradient based Reinforcement Learning}
Policy gradient based methods~\cite{KondaT99} are fundamentally designed to directly learn a parameterized optimal policy $\pi_\theta$. In comparison to value-based methods, policy-based algorithms have the following characteristics: firstly, these methods can be flexibly applied to continuous action spaces; secondly, they can directly obtain a stochastic policy $\pi_\theta(\cdot|s)$. When parameterizing the policy with a neural network, a typical approach is to adjust the parameters in the direction of increasing the cumulative reward: $\theta\leftarrow \theta+\alpha \nabla_{\theta}V^{\pi_\theta}(s)$. However, the gradient is also subject to the unknown effects of policy changes on the state distribution.
In \cite{sutton2018reinforcement}, researchers derived a solution based on policy gradients that does not involve the state distribution:
\begin{equation}
    \begin{aligned}
	\nabla_{\theta}V&^{\pi_\theta}(s)=
\mathbb{E}_{s\sim\mu^{\pi_\theta}(\cdot),a\sim\pi_\theta(\cdot|s)}\left[\nabla_{\theta}\log \pi_\theta(a|s)\cdot Q^{\pi_\theta}(s,a) \right],
    \end{aligned}
\label{sec1:pg}
\end{equation}
where $\mu^{\pi_\theta}$ is the state occupancy measure under policy $\pi_\theta$~\cite{DBLP:conf/nips/HoE16}, and $\nabla_{\theta}\log \pi_\theta(a|s)$ is the policy's score evaluation. When the policy is deterministic, and the action space is continuous, we can further obtain the Deterministic Policy Gradient (DPG) theorem~\cite{SilverLHDWR14}:
\begin{equation}
\begin{aligned}
	\nabla_{\theta}V&^{\pi_\theta}(s)=\mathbb{E}_{s\sim\mu^{\pi_\theta}(\cdot)}\left[\nabla_{\theta}\pi_\theta(a|s)\cdot\nabla_{a}Q^{\pi_\theta}(s,a)|_{a=\pi_\theta(s)} \right].
 \end{aligned}
 \label{sec1:dpg}
\end{equation}
One of the most classical applications of policy gradients is the REINFORCE algorithm\cite{Williams92}, which uses the cumulative reward $R_t=\sum^T_{t'=t}\gamma^{t'-t}r_{t'}$ as an estimate for $Q^{\pi_\theta}(s_t, a_t)$ to update the policy parameters.

\begin{figure*}
	\centering
	\includegraphics[width=0.6\textwidth]{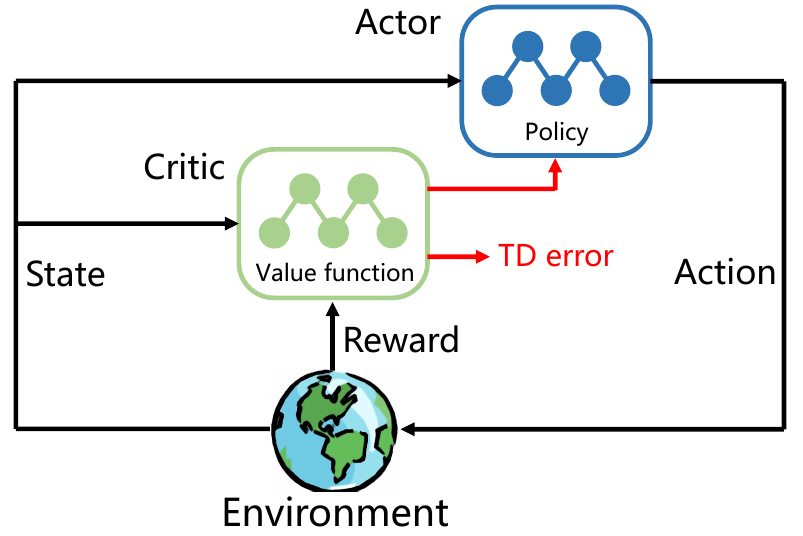}
	\caption{Framework of Actor-Critic based Methods.}
	\label{sec1:2}
\end{figure*}

\subsubsection{Actor-Critic based Reinforcement Learning}
The traditional Actor-Critic method consists of two components: the Actor, which adjusts the parameters $\theta$ of the policy $\pi_\theta$; and the Critic, which adjusts the parameters $w$ of the state-action value function $Q^{\pi_\theta}_w$. Based on these two components, the corresponding update methods for the Actor and Critic can be obtained as follows:
\begin{equation}
    \begin{aligned}
	\theta\leftarrow& \theta + \alpha_{\theta}Q_w(s,a)\nabla_{\theta}\log\pi_\theta(s,a),  \\
    w\leftarrow& w + \alpha_w (r+\gamma Q_w(s', a')- Q_w(s,a))\nabla_w Q_w(s,a).
\end{aligned}
\label{sec1:actor}
\end{equation}

The Actor-Critic method combines policy gradient methods with value function approximation. The Actor probabilistically selects actions, while the Critic, based on the action chosen by the Actor and the current state, evaluates the score of that action. The Actor then modifies the probability of choosing actions based on the Critic's score. The advantage of such methods is that they allow for single-step updates, yielding low-variance solutions in continuous action spaces~\cite{0001BHMMKF17,haarnoja2018soft}. However, the drawback is that, during the initial stages of learning, there is significant fluctuation due to the imprecise estimation by the Critic. Figure~\ref{sec1:2} illustrates the architecture of the Actor-Critic method, and for more details on SARL methods, refer to the review~\cite{li2017deep}.

\subsection{Multi-agent Reinforcement Learning}
\subsubsection{Multi-agent Systems}
The Multi-agent system (MAS) evolved from Distributed Artificial Intelligence (DAI)~\cite{weiss1999multiagent}. The research aim is to address real-world problems that are large-scale, complex, real-time, and involve uncertainty. Modeling such problems with a single intelligent agent is often inefficient and contradictory to real-world conditions. MASs exhibit autonomy, distribution, coordination, self-organization, reasoning, and learning capabilities. Developed since the 1970s, research on MASs encompasses both building individual agents' technologies, such as modeling, reasoning, learning, and planning, and technologies for coordinating the operation of multiple agents, such as interaction, coordination, cooperation, negotiation, scheduling, and conflict resolution. MASs find rapid and extensive applications in areas such as intelligent robotics, traffic control, distributed decision-making, software development, and gaming, becoming a tool for analyzing and simulating complex systems~\cite{dorri2018multi}.

\begin{figure*}
\centering
\includegraphics[width=1\linewidth]{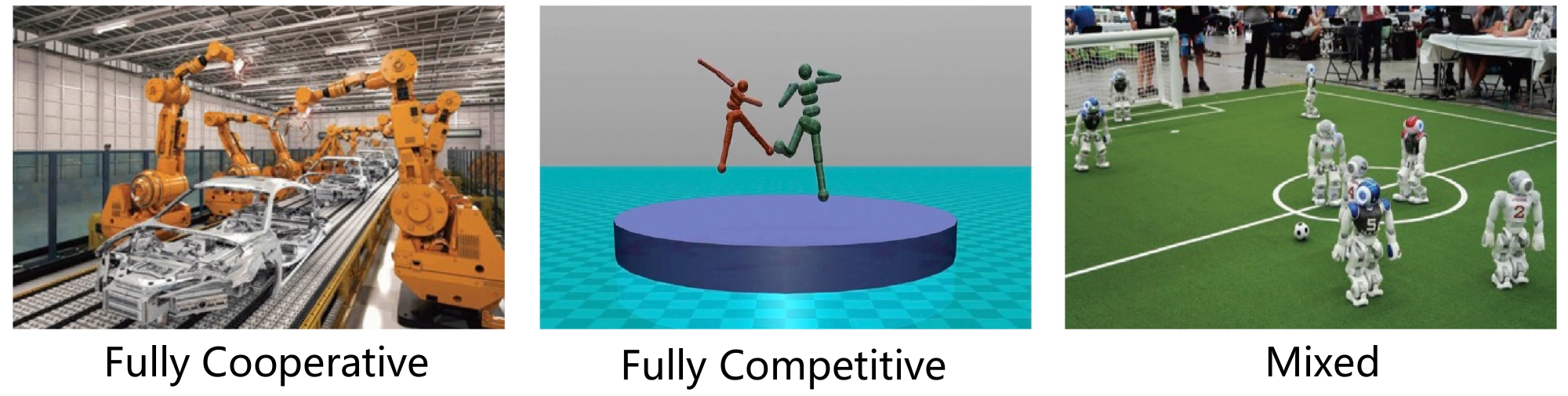}
\caption{Three Common Settings of Multi-agent Systems.}\label{ch17:3mas}
\end{figure*}

Depending on task characteristics, MASs can generally be classified into three settings: Fully Cooperative, Fully Competitive, and Mixed. In a Fully Cooperative setting, agents share a common goal, collaborating to accomplish specific tasks, such as the assembly of machine systems. In a Fully Competitive setting, agents' goals conflict, where one agent's gain results in another's loss, like in a sumo wrestling match. Mixed settings involve agents forming multiple groups, with cooperation within groups and competition between them, resembling a soccer match where teammates cooperate while teams compete.

In MASs, different settings lead to variations in policy optimization goals and learned policies. In cooperative settings, agents must consider teammates' policies, aiming for optimal coordination to avoid interference. In competitive settings, agents consider opponents' policies, minimizing opponents' gains while maximizing their own returns, optimizing their policies accordingly.

\subsubsection{Introduction to Game Theory}
Game theory~\cite{fudenberg1991game,owen2013game}, a mathematical theory and method for analyzing strategic interactions among rational agents, was initially widely applied in economics and has expanded to cover various disciplines, including sociology, politics, psychology, and computer science. In MASs, the interaction and decision-making processes among individuals can be modeled as a game, providing a framework to analyze strategic choices.
Game theory typically models problems as normal-form games, defined as follows:
\begin{definition}[Normal-form Games]
A finite, $n$-player normal-form game consists of a triple $(\mathcal N,\mathcal A, \bm{Q})$, where:
\begin{itemize}
\item $\mathcal N$ is a finite set of agents, with a size of $n$, each indexed by $i\in{1,\cdots,n}$;
\item $\mathcal A=\mathcal A_1\times\cdots\times \mathcal A_n$ represents the joint action space, where each $\mathcal A_i$ is the action set for agent $i$, and $\bm{a}=(a_1,a_2,\cdots,a_n)\in \mathcal A$ is called an action profile.
\item $\bm{Q}=(Q_1,\cdots,Q_n)$, where each component $Q_i:\mathcal A\mapsto \mathbb{R}$ is the utility function of agent $i$.
\end{itemize}
\end{definition}

Although game theory commonly uses the symbol $u$ to denote the utility function, its meaning is equivalent to the action-value function in the case of a single state and 1-horizon MDP. Hence, we replace $u$ with the symbol $Q$ for consistency of notation. We also do not distinguish "policy" and "strategy" in the paper. For a two-player normal-form game where both agents are in a fully competitive scenario, it can be modeled as a zero-sum game:
\begin{definition}[Two-Player Zero-Sum Game]\label{tpzsg}
A game is a two-player zero-sum game when the utility functions satisfy $\forall\bm{a},Q_1(\bm{a})+Q_2(\bm{a})\equiv 0$.
\end{definition}

In a normal-form game, if an agent takes a single action, it is considered a pure policy. When each agent takes a single action, the resulting action profile $\bm{a}$ is known as a pure policy profile. More generally, an agent's policy is defined as follows:
\begin{definition}[Policy and Policy Profile]
In a normal-form game $(\mathcal N,\mathcal A,\bm{Q})$, the policy of agent $i$ is a probability distribution defined on its feasible action space $\mathcal A_i$, denoted as $\pi_i:\mathcal A_i\mapsto[0,1]$. The Cartesian product of all agents' policies, denoted as $\bm{\pi}=\prod_{i=1}^n \pi_i$, is a (mixed) policy profile.
\end{definition}

The symbol $\pi_i$, commonly used in RL, is employed here instead of the often-used symbol $\sigma_i$ in game theory to represent a probability distribution on the feasible action space. Additionally, in game theory, $-i$ usually denotes the set of all players except player $i$. Consequently, the expected utility of agent $i$ under a mixed policy profile $\bm{\pi}=(\pi_i,\bm\pi_{-i})$ is given by
\begin{equation*}
Q_i(\pi_i,\bm \pi_{-i})=\sum_{\bm{a}\in \mathcal A}Q_i(\bm{a})\prod_{i=1}^n \pi_i(a_i).
\end{equation*}

The form of the expected utility shows that, for agent $i$, the expected utility functions of a two-person constant-sum game and a two-player zero-sum game only differ by a constant $c$. Thus, under the same solution concept, there is no difference in the policies between the two types of games. For rational agents, the goal is often to maximize their utility, meaning maximizing their expected returns. Therefore, we need to introduce the definition of best response.

\begin{definition}[Best Response]\label{def_br}
A policy $\pi^\star_i$ for agent $i$ is a best response to the policy profile $\bm{\pi}{-i}$ if
\begin{equation*}
\forall \pi^\prime_i,Q_i(\pi_i^\star,\bm{\pi}{-i})\ge Q_i(\pi_i^\prime,\bm{\pi}{-i}),
\end{equation*}
and the set of best responses to $\bm{\pi}_{-i}$ is denoted as $BR(\bm{\pi}_{-i})$.
\end{definition}

The best response intuitively means that, given the policies of other agents $\bm{\pi}_{-i}$, the agent $i$'s policy $\pi^\star_i$ maximizes its utility. Thus, if other agents' policies are fixed, we can use RL to find the best response. Based on best responses, we can straightforwardly introduce the definition of Nash Equilibrium.

\begin{definition}[Nash Equilibrium]\label{nash_eq}
In a game with $n$ agents, if a policy profile $\bm{\pi}=(\pi_1,\cdots,\pi_n)$ satisfies $\forall i\in{1,\cdots,n},\pi_i\in BR(\bm{\pi}_{-i})$, then $\bm{\pi}$ is a Nash Equilibrium.
\end{definition}

According to the above definition, in a Nash Equilibrium, unilateral deviations by individual agents do not increase their own utility, indicating that each rational agent has no incentive to deviate unilaterally from the equilibrium. Besides Nash Equilibrium, there are many other solution concepts in game theory, such as maximin policy and minimax policy.

\begin{definition} [Maxmin policy]
The maxmin policy for agent $i$ is $\mathop{\arg\max}_{\pi_i}\min_{\bm{\pi}_{-i}}Q_i(\pi_i,\bm{\pi}_{-i})$, and its corresponding maxmin value is $\mathop{\max}_{\pi_i}\min_{\bm\pi_{-i}}u_i(\pi_i,\bm{\pi}_{-i})$.
\end{definition}

The maxmin policy maximizes the agent $i$'s utility in the worst-case scenario (against the most malicious opponent). Therefore, adopting the maxmin policy ensures that the agent's utility is not lower than the maxmin value. When making decisions solely to maximize their own returns without making any assumptions about other agents, the maxmin policy is a suitable and conservative choice. The "dual" policy to the maxmin policy is the minimax policy.

\begin{definition}[Minmax policy]
The minimax policy for other agents $-i$ is $\mathop{\arg\min}_{\bm{\pi}_{-i}}\max_{\pi_{i}}Q_i(\pi_i,\bm{\pi}_{-i})$, and the corresponding minimax value is $\mathop{\min}_{\bm{\pi}_{-i}}\max_{\pi_{i}}Q_i(\pi_i,\bm{\pi}_{-i})$.
\end{definition}

These two policies are often applied in constant-sum games. However, in cooperative settings, attention is given to how agents in the system choose policies through certain coordination means to achieve mutual benefits. Relevant equilibrium provides a reasonable policy coordination mechanism.

\begin{definition}[Correlated Equilibrium]\label{def_coorqu}
A joint policy $\boldsymbol{\pi}$ is a correlated equilibrium if, for $\forall i\in\mathcal N$ and $\forall a^i \in \mathcal{A}^i$,
\begin{align*}
& \sum_{\boldsymbol{a}^{-i}}\boldsymbol{\pi}(a^{i,}, \boldsymbol{a}^{-i})[Q_i(a^{i,}, \boldsymbol{a}^{-i})-Q_i(a^{i}, \boldsymbol{a}^{-i})] \geq 0,
\end{align*}
where $a^{i,*}$ is the best response to $\boldsymbol{a}^{-i}$.
\end{definition}

Correlated equilibrium describes a situation where, assuming two agents follow a correlated policy distribution, no agent can unilaterally change its current policy to obtain a higher utility. It should be noted that Nash Equilibrium implies that choices made by each agent are independent, i.e., the actions of agents are not correlated. Nash Equilibrium can be viewed as a special case of correlated equilibrium.

The concepts mentioned above assume that agents make decisions simultaneously. However, in some scenarios, there may be a sequence of decisions. In such cases, agents are defined as leaders and followers, where leaders make decisions first, enjoying a first-mover advantage, and followers make decisions afterward.

\begin{definition}[Stackelberg Equilibrium]\label{def_staqu}
Assuming a sequential action scenario where leader $\pi_l$ makes policy actions first, and follower $\pi_f$ makes policy actions later. $Q_l$ and $Q_f$ are the utility functions of the leader and follower, respectively. The Stackelberg Equilibrium $(\pi_l^, \pi_{f}^)$ satisfies the following constraints:
$$Q_l(\pi_{l}^*, \pi_f^*) \geq Q_l(\pi_{l}, \pi_f^*), $$
where
$$\pi_f^*=\arg\max_{\pi_f} Q_f(\pi_l,\pi_f).$$
\end{definition}

\subsubsection{Multi-agent Reinforcement Learning}
\begin{figure*}
\centering
\includegraphics[width=0.8\linewidth]{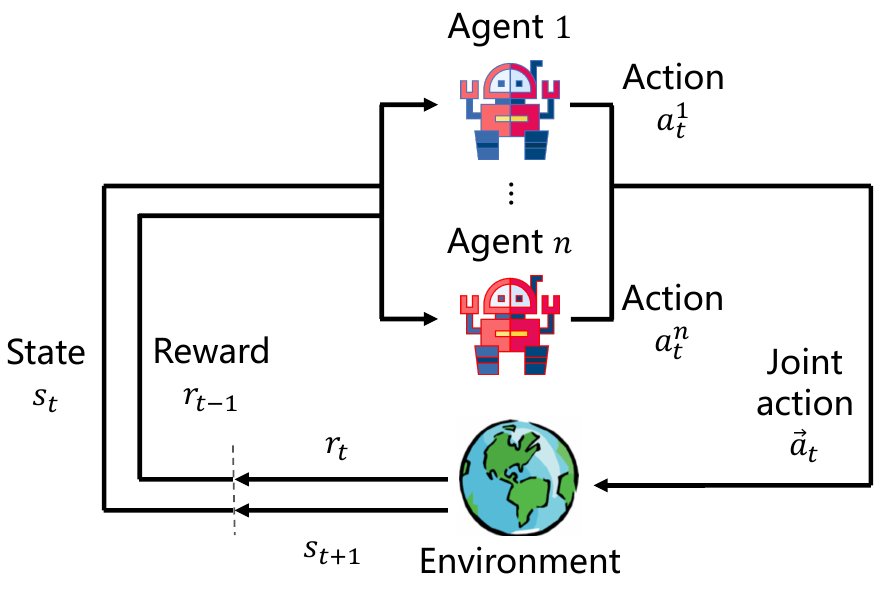}
\caption{Illustration of MARL.}\label{marlfig}
\end{figure*}

Multi-agent Learning (MAL) introduces machine learning techniques into the field of MASs, exploring the design of algorithms for adaptive agent learning and optimization in dynamic environments. The objective of Multi-agent Reinforcement Learning (MARL) (Figure~\ref{marlfig}) is to train multiple agents in a shared environment using RL methods to accomplish given tasks~\cite{hernandez2019survey, gronauer2022multi}.

Differing from modeling the multi-step decision-making process as a Markov Decision Process (MDP) in SARL, MARL is generally modeled as a Stochastic Game.

\begin{definition}[Stochastic Game]\label{sgame}
A stochastic game can typically be represented by a sextuple $\langle\mathcal{N}, \mathcal{S}, \{\mathcal{A}^i\}_{i\in \mathcal{N}},P,$ $\{R^i\}_{i\in \mathcal{N}}, \gamma\rangle$, where:
\begin{itemize}
\item $\mathcal{N}={1,2,...,n}$ represents the set of agents in the system. When $n=1$, the problem degenerates into a single-agent Markov Decision Process, and $n \geq 2$ is a general stochastic game process.
\item $\mathcal{S}$ is the state space shared by all agents in the environment.
\item $\mathcal{A}^i$ is the action space of agent $i$, defining the joint action space $\mathcal{A}:=\mathcal{A}^1 \times \cdots \times \mathcal{A}^n$.
\item $P: \mathcal{S} \times \mathcal{A} \rightarrow \Delta (\mathcal{S})$ is the state transition function, specifying the probability that the environment transitions from $s \in \mathcal{S}$ to another state $s' \in \mathcal{S}$ at each time step, given a joint action $\bm{a} \in \mathcal{A}$.
\item $R^i: \mathcal{S} \times \mathcal{A} \times \mathcal{N} \rightarrow\mathbb R$ is the reward function for each agent.
\item $\gamma\in[0,1]$ is the discount factor.
\end{itemize}
\end{definition}

At each time step, agent $i \in \mathcal{N}$ is in state $s$, selects action $a^i \in \mathcal{A}^i$, forms the joint action $\boldsymbol{a}=\langle a^1, \dots, a^n \rangle$, and executes it. The environment transitions to the next state $s'\sim P( \cdot \mid s, \boldsymbol{a})$, and agent $i$ receives its own reward $R^i(s,\boldsymbol{a})$. Each agent optimizes its policy function $\pi_i :\mathcal{S} \rightarrow \Delta (\mathcal{A}^i)$ to maximize its expected cumulative reward, expressed in the form of a state value function:
\begin{equation}
    \begin{aligned}
    \max_{\pi_i}V^i_{\boldsymbol{\pi}}(s):= &\mathbb{E} [ \sum_{t\geq0}\gamma^t R^i(s_t,\boldsymbol{a}_t) | a_t^i \sim \pi_i(\cdot\mid s_t), \\ \nonumber&\quad\quad\boldsymbol{a}_t^{-i}\sim \boldsymbol{\pi}_{-i}(\cdot\mid s_t), s_0=s],
    \end{aligned}
\end{equation}
where the symbol $-i$ represents all agents except agent $i$. Unlike in SARL where agents only need to consider their own impact on the environment, in MASs, agents mutually influence each other, jointly make decisions, and simultaneously update policies. When the policies of other agents in the system are fixed, agent $i$ can maximize its own payoff function to find the optimal policy $\pi^{\ast}_i$ relative to the policies of other agents. 
In MARL, rationality and convergence are the primary evaluation metrics for learning algorithms.

\begin{definition}[Rationality]
    In the scenario where opponents use a constant policy, rationality is the ability of the current agent to learn and converge to an optimal policy relative to the opponent's policy.

\end{definition}

\begin{definition}[Convergence]
When other agents also use learning algorithms, convergence is the ability of the current agent to learn and converge to a stable policy. Typically, convergence applies to all agents in the system using the same learning algorithm.    
\end{definition}

From the above discussions, it is evident that in the process of MARL, each agent aims to increase its own utility. The learning goal in this case can be to maximize its own Q-function. Therefore, an intuitive learning approach is to construct an independent $Q$-function for each agent and update it according to the $Q$-learning algorithm:

\begin{equation}
\label{rewardlbf}
  \begin{aligned}
   Q_i(s_t, a_t^i)\leftarrow Q_i(s_t, a_t^i)+ \alpha [r_{t+1}+\gamma \max_{a_{t+1}^i\in\mathcal{A}^i} Q_i(s_{t+1},a^i_{t+1})-Q_i(s_t,a^i_t)].
    \end{aligned}
\end{equation}

Similar to the SARL algorithm, each agent selects the current action using a greedy policy:
 \begin{equation}
    \label{pgsarl}
      \begin{split}
       \pi_{i}(a_t^i|s_t)= 
       \mathbbm{1}\{a^i_t=\arg\max_{a^i_t} Q_i(s_t, a^i_t)\}.
        \end{split}
    \end{equation}

In the above scenario, we assume that each agent can obtain global state information for decision-making. However, constrained by real conditions, individuals in MASs often can only obtain limited local observations. To address this characteristic, we generally model such decision processes as Partially-Observable Stochastic Games (POSG).

\begin{definition}[Partially-Observable Stochastic Games (POSG)]
The POSG is often defined as $\mathcal{M} = \langle \mathcal{N}, \mathcal{S}, \{\mathcal{A}^i\}_{i\in \mathcal{N}}, $ $ P, \{R^i\}_{i\in \mathcal{N}}, \gamma ,\{\Omega^i\}_{i\in \mathcal{N}},$ $ \mathcal{O} \rangle$, POSG includes the first six components consistent with the definition~\ref{sgame}. Additionally, it introduces:
\begin{itemize}
\item $\Omega^i$ is the observation space of agent $i$, and the joint observation space for all agents is $\pmb{\Omega}:= \Omega^1 \times \cdots \times \Omega^{n} $.
\item $\mathcal{O}:\mathcal{S} \rightarrow \Delta (\boldsymbol{\Omega})$ represents the observation function, $\mathcal{O}(\boldsymbol{o}|s)$, indicating the probability function concerning the joint observation $\boldsymbol{o}$ given a state $s$.
\end{itemize}
\end{definition}

In POSG, each agent optimizes its policy $\pi_i :\Omega^i \rightarrow \Delta (\mathcal{A}^i)$ to maximize its return. Although POSG is widely applied in real scenarios, theoretical results prove its problem-solving difficulty is NEXP-hard~\cite{bernstein2002complexity}. Fortunately, recent technologies, including centralized MARL training frameworks, utilizing recurrent neural networks to encode historical information~\cite{hausknecht2015deep}, and agent communication~\cite{zhu2022survey}, have alleviated this issue from different perspectives.

In cooperative tasks, POSG can be further modeled as a Decentralized Partially Observable Markov Decision Process (Dec-POMDP)~\cite{bernstein2002complexity}. The main difference lies in the fact that in Dec-POMDP, the reward functions for each agent are identical. Beyond the challenges of partial observability, MARL faces unique problems compared to SARL. In MARL, the state space becomes larger, and the action space experiences the "curse of dimensionality" problem as the number of agents grows~\cite{hernandez2019survey}. Additionally, issues such as scalability, credit assignment, heterogeneity, and cooperative exploration have hindered the development of MARL~\cite{busoniu2008comprehensive}.

\subsubsection{Multi-Agent Reinforcement Learning Training Paradigms}

\begin{figure*} 
\centering
\includegraphics[width=1\linewidth]{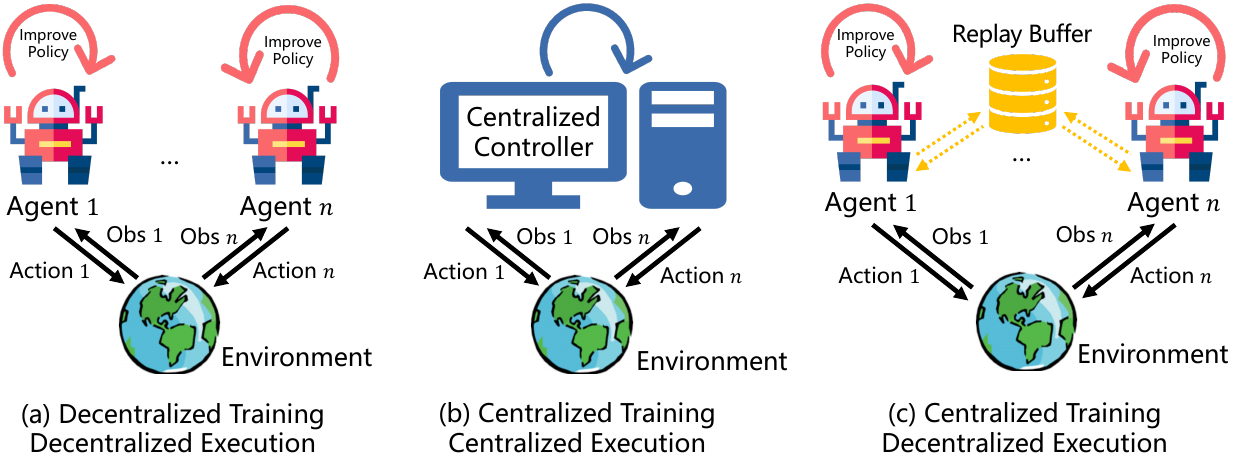}
\caption{Three different training paradigms.}\label{ch17:3tp}
\end{figure*}

In MARL, training is the process of optimizing policies based on acquired experience (states, actions, rewards, etc.), while execution involves agents interacting with the environment by executing actions according to individual or joint policies. Generally, depending on whether agents need information from other agents during policy updates, the training process can be categorized into Centralized Training and Decentralized Training. Correspondingly, based on whether external information is required during the execution phase, it is categorized into Centralized Execution and Decentralized Execution. Combining these phases, MARL encompasses three paradigms: Decentralized Training Decentralized Execution (DTDE), Centralized Training Centralized Execution (CTCE), and Centralized Training Decentralized Execution (CTDE), as illustrated in Figure~\ref{ch17:3tp}.

\begin{definition}[Decentralized Training Decentralized Execution (DTDE)]
In the DTDE framework, each agent independently updates its policy and executes using only its local information, without any information exchange. The policy is represented as $\pi_i:\Omega^i\rightarrow \Delta(\mathcal{A}^i)$.    
\end{definition}

IQL \cite{tan1993multi} is a typical algorithm based on DTDE, known for its scalability as it only considers the individual agent. However, due to the lack of consideration for information from other agents, agents often operate in a non-stationary environment, as defined in Section 2.1. To address the inefficiency brought by decentralized learning, optimistic and delayed optimization assign greater update weights to better $Q$ values, alleviating the non-stationarity issue. Techniques like recurrent neural networks can also mitigate these problems.

When updating policies using DQN, direct usage of data from experience replay buffer exacerbates non-stationarity. Importance sampling can be employed to alleviate this issue:
$$\mathcal{L}(\theta_i)=\frac{\boldsymbol{\pi}_{-i}^{t_c}(\boldsymbol{a}^{-i}|\boldsymbol{o}^{-i})}{\boldsymbol{\pi}_{-i}^{t_i}(\boldsymbol{a}^{-i}|\boldsymbol{o}^{-i})}[(y_i^{\rm Q}-Q_i(o^i,a^i;\theta_i))^2],$$
where $\theta_i$ is the $Q$ network parameters for agent $i$, $t_c$ is the current time, $t_i$ is the sample collection time, and $y_i^Q$ is the temporal difference target, calculated similarly to the single-agent case.

\begin{definition}[Centralized Training Centralized Execution (CTCE)]
In the CTCE framework, agents learn a centralized joint policy:
$$\boldsymbol{\pi}:\boldsymbol{\Omega}\rightarrow \Delta(\boldsymbol{\mathcal{A}}),$$    
and exectues accordingly.
\end{definition}

Here, any SARL algorithm can be used to train MASs. However, the complexity of the algorithm grows exponentially with the dimensions of states and actions~\cite{foster2023complexity}. This issue can be addressed through policy or value decomposition.

While decomposition methods alleviate the dimension explosion problem, CTCE struggles to evaluate the mutual influences among agents. Recently, attention has been given to the Centralized Training Decentralized Execution (CTDE) framework, especially in complex scenarios.

\begin{definition}[Centralized Training Decentralized Execution (CTDE)]\label{def_ctde}
In the CTDE framework, during the training phase, agents optimize their local policies with access to information from other agents or even global information:
$\pi_i:\Omega^i\rightarrow \Delta(\mathcal{A}^i).$
During the decentralized execution process, agents make decisions using only their local information.
\end{definition}

CTDE has found widespread applications in MASs, particularly excelling in certain complex scenarios. However, when dealing with heterogeneous MASs, CTDE may face challenges. Skill learning or grouping followed by local CTDE training are proposed solutions. Additionally, research has discussed centralized and decentralized aspects from various perspectives~\cite{lyu2023centralized, zhou2023centralized}.

\subsubsection{Challenges and Difficulties in Multi-Agent Reinforcement Learning}

Compared to SARL, real-world scenarios are often better modeled as MARL. However, the presence of multiple agents simultaneously updating their policies introduces more challenges. This section discusses key challenges in MARL, including non-stationarity, scalability, partial observability, and current solutions.

In a single-agent system, the agent only considers its own interaction with the environment, resulting in a fixed transition of the environment. However, in MASs, simultaneous policy updates by multiple agents lead to a dynamic target process, termed non-stationarity.

\begin{definition}[Non-Stationarity Problem]

In multi-agent systems, a particular agent $i$ faces a dynamic target process, where for any $\pi_i \neq \overline{\pi}_i$:

$$ P(s'| s, \boldsymbol{a}, \pi_i,\boldsymbol{\pi}_{-i}) \neq P(s'| s, \boldsymbol{a}, \bar \pi_i,\boldsymbol{\pi}_{-i}). $$
\end{definition}
Non-stationarity indicates that, with simultaneous learning and updates by multiple agents, learning no longer adheres to Markovian properties. This issue is more severe in updates of $Q$ values, particularly methods relying on experience replay, where sampling joint actions in multi-agent environments is challenging. To address non-stationarity, solutions involve modeling opponents (teammates), importance sampling of replay data, centralized training, or incorporating meta-RL considering updates from other teammates.

\begin{definition}[Scalability Issue]
    To tackle non-stationarity, multi-agent systems often need to consider the joint actions of all agents in the environment, leading to an exponential rise in joint actions with an increasing number of agents.
\end{definition}

In real-world scenarios, the number of agents in MASs is often substantial, such as in autonomous driving environments where there may be hundreds or even thousands of agents. Scalability becomes essential but is also a highly challenging aspect. Current methods to address scalability include parameter sharing, where homogeneous agents share neural networks, and heterogeneous agents undergo independent training~\cite{christianos2021scaling}. Alternatively, techniques such as transfer learning or curriculum learning involve training initially with fewer agents and progressively scaling up to environments with a larger number of agents~\cite{da2019survey}. 

Due to sensor limitations and other factors, agents often struggle to obtain global states and can only access partial information. MARL is often modeled as Partially Observable Stochastic Games (POSG), where the environment, from an individual agent's perspective, no longer adheres to Markovian properties, posing training challenges.

\begin{figure*}
\centering
\includegraphics[width=1\linewidth]{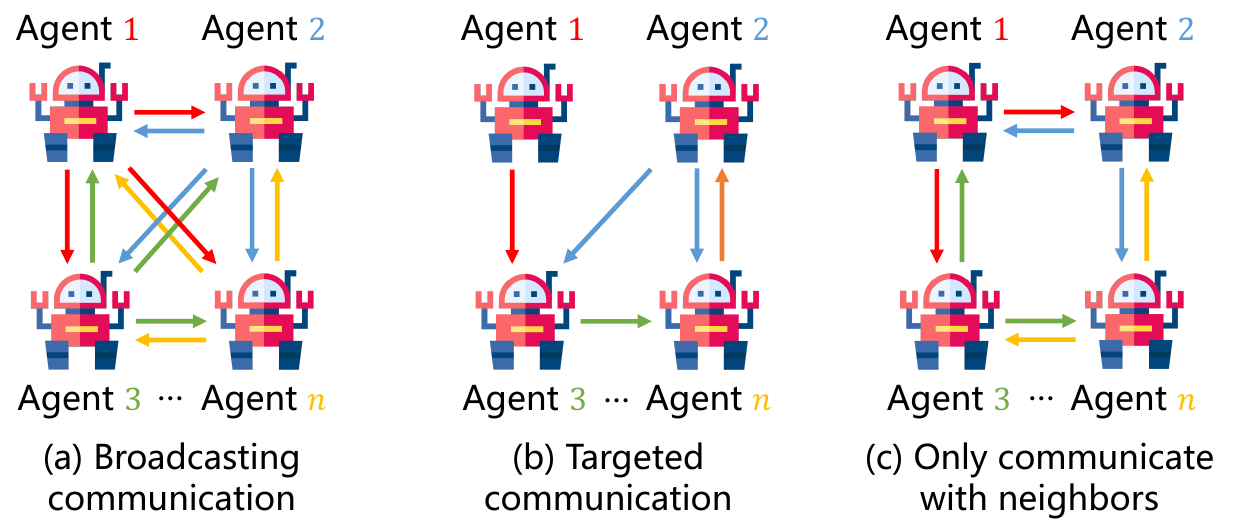}
\caption{Three different communication topologies.}\label{ch17:3com}
\end{figure*}

Various solutions have been proposed, including multi-agent communication, where agents exchange information to alleviate local observation issues. Key questions in communication involve determining who to communicate with, what information to communicate, and when to communicate. Current research explores three communication topologies (Figure \ref{ch17:3com}): broadcasting information to all agents, forming network structures for selective communication, or communicating with a subset of neighbors. Policies for handling these questions include direct exchange of local information, preprocessing information before transmission, and optimizing communication timing.
In conclusion, the challenges of non-stationarity, scalability, and partial observability in MARL demand innovative solutions to ensure effective learning and decision-making in complex, real-world scenarios.

\section{Cooperative Multi-agent Reinforcement Learning in Classical Environments
}
\subsection{Concept of Multi-agent Cooperation
}\label{sec-MArldf}

Previously, we mentioned that MASs can be categorized into fully cooperative, fully competitive, and mixed settings based on task characteristics. Among these settings, competitive scenarios are generally solved by modeling them as zero-sum games, while mixed scenarios involve a combination of cooperation and competition. This section focuses on cooperative MARL algorithms, which have been extensively researched.

\begin{definition}[Fully Cooperative]\label{def_fullc}
In MARL, fully cooperative refers to all agents sharing a common reward function, satisfying the condition: 
$$R^1=R^2= \cdots =R^n=R.$$
\end{definition}

In cooperative MARL, where agents share a common reward function, it is evident that agents have identical returns. With a centralized controller, the learning process can degrade into a Markov Decision Process, and the action space of agents becomes a joint action space for a stochastic game. The optimization objective is as follows:

\begin{equation}
    \begin{aligned}
        Q(s_t, \boldsymbol{a}_t)\leftarrow 
        Q(s_t, \boldsymbol{a}_t)+\alpha [r_{t}+\gamma \max_{\boldsymbol{a}_{t+1}\in\boldsymbol{\mathcal{A}}} Q(s_{t+1},\boldsymbol{a}_{t+1})-Q(s_t,\boldsymbol{a}_t)].
    \end{aligned}
\end{equation}

However, in real-world scenarios, this objective is challenging for multi-agent decision-making because agents often independently make decisions. Concretely, agents may choose actions belong to different optimal joint actions, due to decentralized execution. Ignoring inter-agent cooperation behavior~\cite{lauer2000algorithm}, early distributed Q-learning algorithms assumed a single optimal equilibrium point, allowing for the direct use of the above formula, where each agents only update its local action-value function $Q_i(s, a^i)$.

\begin{definition}[Credit Assignment Problem]\label{def_fullca}
In fully cooperative MARL, where agents share global rewards, accurately evaluating the contribution of an agent's action to the entire system becomes a challenge.
\end{definition}

\begin{definition}[Coordination-Free Multi-Agent Cooperation Method]\label{def_coordinationfree}
Assuming each agent's local action-value function is $Q_i(s_t,a^i_t)$, its update is as follows:
 \begin{equation}
    \begin{aligned}
        Q_i(s_t, a_t^i)\leftarrow 
        \max\{Q_i(s_t, {a}^i_t),
        \ r_{t}+\gamma \max_{a^i_{t+1}\in\mathcal{A}^i} Q_i(s_{t+1},{a}^i_{t+1})\}.
    \end{aligned}
\end{equation}
\end{definition}

 Coordination-free approach ignores poorly rewarded actions. However, consistently choosing the maximum operation may lead to overestimation issues, and it often lacks scalability due to the calculation based on the joint action function~\cite{claus1998dynamics}.

\begin{definition}[Joint Action Learnng]\label{def_jal}
Assuming agents are in a globally observable environment, observing the states and actions of all teammates, each agent's policy is $\pi_i$, and there exists a global value function $Q(s, \boldsymbol{a})$, then for agent $i$, its local action value can be assessed as:
$$Q_i(s_t, a_i) =: \sum_{\boldsymbol{a}_{-i} \in \boldsymbol{\mathcal{A}}_{-i}}Q(\langle a_i,\boldsymbol{a}_{-i}\rangle)\prod_{j \neq i} {\pi_j}(a_j).$$
\end{definition}

Simultaneously, in special scenarios, multi-agent cooperation can be indirectly achieved through role assignment, coordination graphs, multi-agent communication, etc. Additionally, distributed systems often obtain only local observations. We generally model cooperative MARL as a partially observable stochastic game (POSG), a special case of Dec-POMDP. In Dec-POMDPs, each agent shares the same reward function. We observe that when all agents share a reward function and return, some agents can receive rewards without making a substantial contribution to the system, leading to the credit assignment problem in the MAS~\cite{foerster2018counterfactual, wang2021towards}.

\subsection{Typical Cooperative MARL Algorithms}
\subsubsection{Policy-Gradient-Based MARL Methods}

In cooperative MARL, where rewards and returns among agents are identical, all agents possess the same value function. We denote the global action value function as \(Q_{\boldsymbol{\pi}}(s, \boldsymbol{a})\) and the state value function as \(V_{\boldsymbol{\pi}}(s)\). It's worth noting that, in MARL, the action value function and state value function of an agent depend on the policies of all agents:

$$\pi_1(a^1|o^1, \theta^1), \pi_2(a^2|o^2, \theta^2), \cdots, \pi_n(a^n|o^n, \theta^n).$$

Here, \(\theta^i\) represents the policy parameters of agent \(i\), and the differences in policies among agents mainly manifest in the differences in policy parameters. In the process of policy learning, our objective is to optimize policy parameters to maximize the objective function:

$$ J(\theta^1, \cdots, \theta^n) = {E}_s[V_{\boldsymbol{\pi}}(s)].$$

All agents share a common goal, optimizing their policy parameters \(\theta^i\) to maximize the objective function \(J\). Thus, the optimization objective can be expressed as follows:

$$ \max_{\theta^1, \cdots, \theta^n}J(\theta^1, \cdots, \theta^n).$$

For a specific agent, by performing gradient ascent, we maximize the objective function:

$$\theta^i \leftarrow \theta^i + \alpha^i \nabla_{\theta^i}J(\theta^1, \cdots, \theta^n).$$

Here, \(\alpha^i\) is the learning rate, and the stopping criterion is the convergence of the objective function. In the above equation, direct computation of the gradient is not feasible, and a value network is commonly employed to approximate the policy gradient.

\begin{theorem}[Policy Gradient Theorem of Cooperative MARL]\label{the_cmarlpg}
Suppose there exists a baseline function \(b\) independent of joint actions. Then, cooperative MARL has the following policy gradient:

\begin{equation}
    \begin{aligned}
        \nabla_{\theta^i}J(\theta^1, \cdots, \theta^n) =  \mathbb{E}[\nabla_{\theta^i}\log\pi_i(a^i|o^i,\theta^i)\cdot (Q_{\boldsymbol\pi}(s,\boldsymbol{a})-b)].
    \end{aligned}
\end{equation}

Here, the joint actions is sampled from the joint action distribution:

$$ \boldsymbol{\pi}(\boldsymbol{a}|\boldsymbol{o},\boldsymbol{\theta})= \pi_1(a^1|o^1;\theta^1)\times \cdots  \times \pi_n(a^n|o^n;\theta^n) $$
\end{theorem}

\begin{figure*}
\centering
\includegraphics[width=0.45\linewidth]{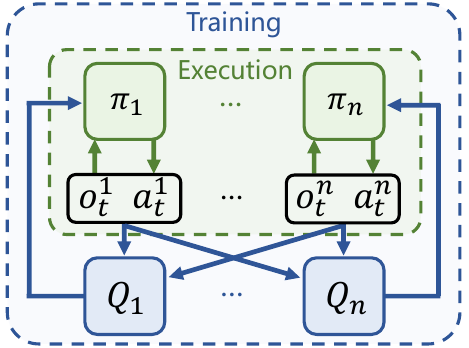}
\caption{Framework of MADDPG~\cite{maddpg}.}\label{maddpgfig}
\end{figure*}

Here, we focus on the Multi-Agent Deep Deterministic Policy Gradient (MADDPG) algorithm \cite{maddpg} (Figure~\ref{maddpgfig}). The algorithm's pseudocode is provided in Algorithm~\ref{maddpgcode}. One one hand, each agent corresponds to a distributed deterministic policy, the Actor module \(\pi_i(o^i;\theta^i)\), and decisions are made by agents based on their respective Actors. On the other hand, each agent is equipped with a centralized Critic network \(Q_i(s,\boldsymbol{a};\phi^i)\) to guide the updating and optimization of the Actor network. Specifically, MADDPG additionally maintains a target policy \(\pi_i'(o^i;\theta^i_{-})\) for delayed updates to calculate the temporal difference target value for the Critic:
\begin{equation}
    \begin{aligned}
    &y^Q_i = r + \gamma Q_i(s', \boldsymbol{a}'|\phi^i)|_{{a'}^j=\pi_j'({o'}^j;\theta^j_{-})},\\
    &\mathcal{L}(\phi^i) = \frac{1}{2}(y_i^Q-Q_i(s,\boldsymbol{a};\phi^i))^2.
    \end{aligned}
    \label{maddpg_critic}
\end{equation}
Furthermore, we use the Critic to guide the Actor with the following update:
\begin{equation}
    \begin{aligned}
        \nabla_{\theta^i}J(\theta^i) = \nabla_{\theta^i}\pi_i(o^i;\theta^i)\nabla_{a^i} Q_i(s, \boldsymbol{a};\phi^i)|_{a^j=\pi_j(o^j|\theta^j)}
    \end{aligned}
    \label{maddpg_actor}
\end{equation}

\begin{algorithm}[h!]\caption{MADDPG Algorithm}\label{maddpgcode}
    \begin{algorithmic}[1]
      \FOR{$episode=1$ \TO $M$}
          \STATE Initialize random exploration noise $\mathcal{N}$
          \STATE Obtain initial state $s$
          \FOR{$t=1$ \TO maximum trajectory length}
            \STATE Each agent explores based on the current policy, obtaining actions: $a^i=\pi_i(o^i;\theta^i)+\mathcal{N}_t$
            \STATE All agents execute joint actions $\boldsymbol{a}=(a^1,\cdots,a^n)$, obtaining rewards $r$ and new state $s'$
            \STATE Store $(s,\boldsymbol{a},r,s')$ in the experience pool $\mathcal{D}$
            \STATE $s \leftarrow s'$
            \FOR{agent $i=1$ \TO $n$}
                \STATE Sample samples $\{s_j,\boldsymbol{a}_j,r_j,s'_{j}\}_{j=1}^{bs}$ from experience pool $\mathcal{D}$
                \STATE Optimize Critic network according to Equation~\ref{maddpg_critic}
                \STATE Optimize Actor network according to Equation~\ref{maddpg_actor}
            \ENDFOR
            \STATE Update target neural networks: $\theta^i_{-}=\tau \theta^i+ (1-\tau) \theta^i_{-}$
          \ENDFOR
      \ENDFOR
    \end{algorithmic}
\end{algorithm}

MADDPG is a typical algorithm based on the CTDE framework. Training the critic network requires access to the environment's global state information. In environments where global state information is unavailable, usually only the observation information \(o_i\) of other agents is accessible. As the number of agents increases, training the Critic network becomes more challenging. Techniques such as attention mechanisms \cite{iqbal2019actor} can be employed for information aggregation to alleviate the computational cost arising from changes in quantity. Additionally, MADDPG can handle competitive settings. However, due to privacy concerns, it may not be possible to obtain the actions of other agents. In such cases, opponent modeling \cite{albrecht2018autonomous} can be used to estimate the action information of other agents.

MADDPG in cooperative MARL tasks does not explicitly consider the credit assignment problem. To explicitly model the contributions of each agent in cooperative MASs (or the rewards they should receive), the Counterfactual Multi-Agent Policy Gradients (COMA) algorithm \cite{foerster2018counterfactual} is widely used. First, the counterfactual baseline is defined:
\begin{definition}[Counterfactual Baseline for Cooperative Multi-Agent Systems]
    In cooperative MARL, assuming the existence of a global state-action value function \(Q(s, \boldsymbol{a})\), when the policies of other agents are fixed, the advantage function for agent \(i\) at the current action is defined as:
\[A_i(s,\boldsymbol{a})= Q(s,\boldsymbol{a})- \sum\limits_{a'^i}\pi^i(a'^i|o^i)Q(s,\langle\boldsymbol{a}^{-i},a'^i\rangle).\]
\end{definition}

In the above definition, \(A_i(s,\boldsymbol{a})\) calculates the baseline function using a centralized Critic for each agent \(i\). This baseline function is then used to compute its policy gradient:
\begin{definition}[Policy Gradient based on COMA]
    For a MARL algorithm based on the Actor-Critic framework and considering \(TD(1)\) Critic optimization, the policy gradient is given by:
\[\nabla J = \mathbb{E}\left[\sum_{i=1}^n\nabla_{\theta^i} \log \pi_i(a^i|o^i)A_i(s,\boldsymbol{a})\right].\]
\end{definition}
\subsubsection{Value Decomposition MARL Methods}
While CTDE-based multi-agent policy gradient methods show promising performance in some application scenarios, and methods based on attention mechanisms, such as MAAC \cite{iqbal2019actor}, can alleviate the curse of dimensionality to some extent, and methods based on counterfactual baselines can address the credit assignment problem to a certain extent. However, in complex cooperative scenarios, such as micro-management in StarCraft \cite{pymarl}, these methods are often inefficient. On the contrary, MARL methods based on value decomposition have shown better performance \cite{gorsane2022towards}. In the following discussion, we will change the input of the value function from observation \(o_t^i\) to trajectory \(\tau_t^i=(o_1, a_1,...,a_{t-1}, o_t)\in\mathcal{T}^i\), thus partially alleviating the problems caused by partial observability. It is worth noting that this technique can also be used in policy-gradient-based methods mentioned above to improve coordination performance. Most value decomposition methods are based on the Individual-Global-Max (IGM) principle \cite{owen2013game}.

\begin{theorem}[Individual-Global-Max Principle]
    For a joint action value function \(Q_{\rm tot}:\boldsymbol{\mathcal T} \times \boldsymbol{\mathcal{A}}\) and local individual action value functions \(\{Q_i:\mathcal{T}^i \times \mathcal{A}^i \rightarrow \mathbb{R}\}_{i=1}^n\), if the following equation holds:

\[ \arg\max\limits_{\boldsymbol{a}} Q_{\rm tot}(\boldsymbol{\tau}, \boldsymbol{a})=
\left(
    \begin{array}{lc}
       & \arg \max\limits_{a^1} Q_{1}(\tau^1, a^1)  \\
       & \vdots   \\ 
       & \arg \max\limits_{a^n} Q_{n}(\tau^n, a^n)
    \end{array}
\right),\]

then, \(Q_{\rm tot}\) and \(\{Q_i\}_{i=1}^n\) satisfy the Individual-Global-Max principle.
\end{theorem}

Three typical value decomposition methods that satisfy the IGM principle are as follows. VDN \cite{vdn} represents the joint Q function as the sum of local Q functions:
\begin{equation}
     Q_{\rm tot}^{\mathrm{VDN}}(\boldsymbol{\tau}, \boldsymbol{a})=\sum_{i=1}^{n} Q_{i}\left(\tau^{i}, a^{i}\right).
\end{equation}

Although this value decomposition method satisfies IGM, its network expressiveness is often insufficient and it performs poorly in complex scenarios. QMIX \cite{qmix} (see Figure~\ref{qmixflow}) extends VDN by introducing a mixing network that combines the Q values of each agent with the global state to calculate the global utility and perform credit assignment. To ensure IGM, the mixing network takes the individual Q values and the state as inputs, calculates the global Q value, and satisfies certain conditions:
\begin{figure*}
\centering
\includegraphics[width=0.8\linewidth]{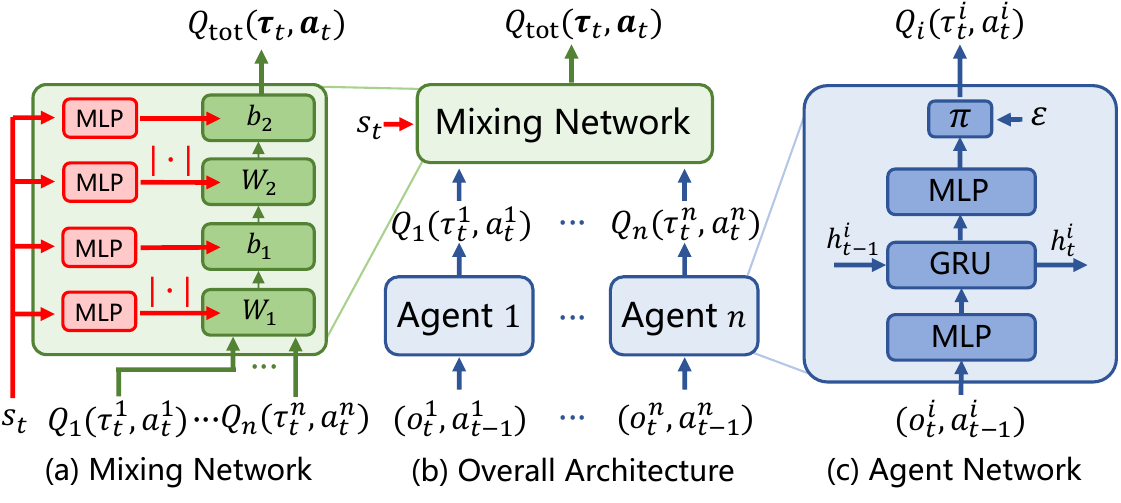}
\caption{Framework of QMIX.}\label{qmixflow}
\end{figure*}

\begin{equation}
\begin{aligned}
    Q_{\rm tot}^{\mathrm{QMIX}}(\boldsymbol{\tau}, \boldsymbol{a}) = &Mixing(s, Q_1(\tau^1, a^1), ..., Q_n(\tau^n, a^n)),\\
    \forall & i \in \mathcal{N}, \frac{\partial Q_{\rm tot}^{\mathrm{QMIX}}(\boldsymbol{\tau}, \boldsymbol{a})}{\partial Q_{i}\left(\tau^{i}, a^{i}\right)}\geq 0.
 \end{aligned}
 \label{qmix_q}
\end{equation}

During optimization, the temporal difference loss is computed for optimization:
\begin{equation}
\begin{aligned}
    \mathcal{L}_{\text{QMIX}}=(Q_{\text{tot}}^{\text{QMIX}}(\boldsymbol{\tau}, \boldsymbol{a})-y_{\text{tot}})^2,
 \end{aligned}
 \label{qmix_loss}
\end{equation}

where \(y_{\text{tot}}=r+\gamma\max_{\boldsymbol{a}'} Q_{\text{tot}}^{\text{QMIX},-}(\boldsymbol{\tau}', \boldsymbol{a}')\) and \(Q_{\text{tot}}^{\text{QMIX},-}\) is the target Q network. Based on the expressiveness and simplicity of the mixing network, QMIX achieves superior results in various environments, and its pseudocode is shown in Algorithm~\ref{qmix}.

\begin{algorithm}[h!]\caption{QMIX Algorithm}\label{qmix}
    \begin{algorithmic}[1]
      \STATE Initialize network parameters $\theta$, including the mixing network, local neural networks for agents, and a hypernetwork.
      \STATE Set learning rate $\alpha$, clear experience replay buffer $D=\{\}$, $t=0$
      \WHILE{training not finished}
          \STATE Set $\text{step}=0$, $s_0$ as the initial state
          \WHILE{$s_t$ not in terminal state and step is less than the maximum trajectory length}
              \FOR{each agent $i$}
                  \STATE $\tau_t^i=\tau_{t-1}^i\cup \{(o_t^i,a_{t-1}^i)\}$
                  \STATE $\epsilon=\text{epsilon\_schedule}(t)$
                  \STATE $a_t^i=\begin{cases}
                  \arg\max_{a^i_t} Q(\tau^i_t, a^i_t)\quad \text{with probability}\,1-\epsilon\\
                  \text{Randint}(1, |\mathcal{A}^i|)\quad\quad \text{with probability}\,\epsilon
                  \end{cases}$
              \ENDFOR
              \STATE Obtain reward function $r_t$ and the next state $s_t$
              \STATE Let $D = D\cup\{ (s_t, \boldsymbol{a}_t, r_t, s_{t+1})\}$
              \STATE $t=t+1$, $\text{step}=\text{step}+1$
          \ENDWHILE
          \IF{$|D|\geq \text{batch\_size}$}
              \STATE  $b \leftarrow $ randomly sample from $D$
              \FOR{each time point $t$ in each trajectory of $b$}
                  \STATE Compute $Q_{\text{tot}}^{\text{QMIX}}$ and $y_{\text{tot}}^{\text
{QMIX}}$ through Equation~\ref{qmix_q}
              \ENDFOR
              \STATE  Compute $\mathcal{L}_{\text{QMIX}}$ through Equation~\ref{qmix_loss} for optimization.
          \ENDIF
          \IF{update time exceeds the designed threshold}
            \STATE Update target networks
          \ENDIF
      \ENDWHILE
    \end{algorithmic}
\end{algorithm}

The conditions satisfied by VDN and QMIX are sufficient but unnecessary conditions for the IGM property, leading to some shortcomings in their expressiveness. To further enhance the expressiveness of neural networks, QPLEX \cite{qplex} proposed a new Duplex Dueling Multi-agent (DDMA) structure:
\begin{equation}
\begin{aligned}
     Q_{\rm tot}^{\mathrm{QPLEX}}(\boldsymbol{\tau}, \boldsymbol{a})=V_{\rm tot}(\boldsymbol{\tau})+A_{\rm tot}(\boldsymbol{\tau}, \boldsymbol{a})
     =
     \sum_{i=1}^{n} Q_{i}\left(\boldsymbol{\tau}, a^{i}\right)+\sum_{i=1}^{n}\left(\lambda^{i}(\boldsymbol{\tau}, \boldsymbol{a})-1\right) A_{i}\left(\boldsymbol{\tau}, a^{i}\right).
     \end{aligned}
\end{equation}
Here, \(\lambda^{i}(\boldsymbol{\tau}, \boldsymbol{a})\) is computed using a multi-head attention mechanism \cite{vaswani2017attention}, obtaining the credit assignment for different advantage functions.
QPLEX, with its expressiveness and network structure, achieves superior results in various environments.

\subsubsection{Integrating Policy Gradient and Value Decomposition in Cooperative MARL}

\begin{figure*}
\centering
\includegraphics[width=0.3\linewidth]{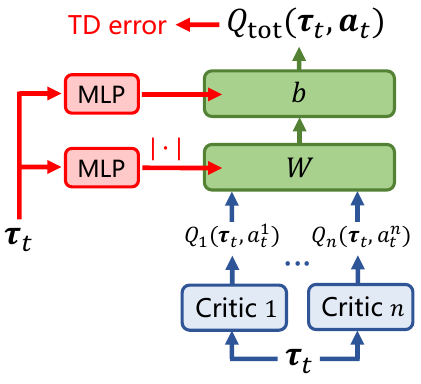}
\caption{Framework of DOP.}\label{dopfigure}
\end{figure*}

In addition to the aforementioned methods based on policy gradient and value decomposition, some works combine the advantages of both to develop algorithms. One such algorithm is Decomposed Off-policy multi-agent Policy gradients (DOP) \cite{wang2020dop} (as shown in Figure~\ref{dopfigure}), which introduces the idea of value decomposition into the multi-agent actor-critic framework and learns a centralized but decomposed critic. This framework decomposes the centralized critic into a weighted linear sum of individual critics with local actions as inputs. This decomposition structure not only achieves scalable learning of critics but also brings several benefits. It realizes feasible off-policy evaluation for stochastic policies, alleviates the centralized-decentralized inconsistency problem, and implicitly learns efficient multi-agent reward allocation. Based on this decomposition, DOP develops efficient off-policy multi-agent decomposed policy gradient methods for both discrete and continuous action spaces.

DOP's specific contributions include: (1) To address the credit assignment problem in MASs and the scalability problem of learning a centralized critic, DOP uses linear value function decomposition to learn a centralized critic, enabling local policy gradients to be derived for each agent based on individual value functions. (2) DOP proposes off-policy MARL algorithms for both stochastic policy gradients and deterministic policy gradients, effectively handling tasks with discrete and continuous action spaces, respectively. (3) Considering the sample efficiency issue of policy gradient algorithms, DOP combines linear value function decomposition with tree-backup techniques, proposing an efficient off-policy critic learning method to improve the sample efficiency of policy learning. In addition to DOP, other work combines the advantages of value decomposition and policy gradient from different perspectives. VDAC \cite{su2021value} directly integrated policy gradient-like methods with value decomposition methods, proposing methods based on linear summation and mixed neural networks. IAC \cite{ma2021modeling}, based on MAAC, further optimizes the credit assignment between agents through value decomposition methods, achieving surprising coordination performance in various environments. FACMAC \cite{peng2021facmac} proposed a decomposable joint policy gradient multi-agent algorithm and proposes the continuous environment multi-agent testing environment MAMuJoCo, FOP \cite{zhang2021fop} designed a value decomposition and gradient decomposition method based on maximum entropy, RIIT \cite{hu2021rethinking} integrated some common techniques in current MARL into the coordination algorithms and proposes a new open-source algorithm framework PyMARL2~\footnote{https://github.com/hijkzzz/pymarl2}.

\subsection{Cooperative Multi-Agent Reinforcement Learning in Classical Environments}

In addition to methods aimed at improving cooperative capabilities discussed in the previous section, researchers have explored cooperative MARL from various perspectives, including efficient exploration, communication, and more. The core content, representative algorithms, and applications or achievements in various research directions in classical environments are summarized in Table~\ref{closeenvironment}.

\begin{table*}
\renewcommand{\arraystretch}{1}
    \centering
    \caption{Research directions of Cooperative MARL in classical environments.}
    \resizebox{0.94\textwidth}{!}{
    \begin{tabularx}{\textwidth}{YZYZ}
    \toprule
    Research Direction & Core Content & Representative Algorithms & Applications and Achievements\\
    \midrule
    Algorithm Framework Design & Utilize multi-agent coordination theories or design neural networks to enhance coordination & VDN~\cite{vdn}, QMIX~\cite{qmix}, QPLEX~\cite{qplex}, MADDPG~\cite{maddpg}, MAPPO~\cite{mappo}, HAPPO~\cite{kuba2021trust}, DOP~\cite{wang2020dop}, MAT~\cite{wen2022multiagent} & Demonstrated significant cooperative potential in various tasks such as SMAC~\cite{pymarl}, GRF~\cite{wen2022multiagent}, showcasing good coordination effects \\
         \midrule
    Cooperative Exploration & Design mechanisms for efficient exploration of the environment to obtain optimal coordination patterns. Simultaneously, collect efficient trajectory experiences to train policies to find optimal solutions & MAVEN~\cite{DBLP:conf/nips/MahajanRSW19}, EITI(EDTI)\cite{wang2019influence}, EMC\cite{zheng2021episodic}, CMAE~\cite{liu2021cooperative}, Uneven~\cite{gupta2021uneven}, SMMAE~\cite{zhang2023self} & Significant improvement in cooperative performance in complex task scenarios, addressing low coordination ability in sparse reward scenarios \\
    \midrule
    Multi-Agent Communication & Design methods to promote information sharing between agents, addressing issues such as partial observability. Focus on when and with which teammate(s) to exchange what type of information & DIAL~\cite{foerster2016learning}, VBC~\cite{vbc}, I2C~\cite{i2c}, TarMAC~\cite{tarmac}, MAIC~\cite{maic}, MASIA~\cite{guan2023efficient} & Effectively enhances cooperative capabilities in scenarios with partial observability or requiring strong coordination \\
\midrule
    Agent Modeling & Develop technologies to endow agents with the ability to infer the actions, goals, and beliefs of other agents in the environment, promoting the improvement of cooperative system capabilities & ToMnet~\cite{rabinowitz2018machine}, OMDDPG~\cite{papoudakis2020variational}, LIAM~\cite{papoudakis2021agent}, LILI~\cite{xie2021learning}, MBOM~\cite{yu2022model}, MACC~\cite{macc} & Significant improvement in cooperative performance in scenarios with environmental non-stationarity due to the presence of other agents, improvement in performance in scenarios with strong interaction and the need for strong coordination \\
\midrule
    Policy Imitation & Agents learn cooperative policies from given trajectories or example samples to accomplish tasks & MAGAIL~\cite{DBLP:conf/nips/SongRSE18}, MA-AIRL~\cite{yu2019multi}, CoDAIL~\cite{liu2019multi}, DM$^2$\cite{wang2022dm} & Achievement of policy learning solely from example data \\
\midrule

    Model-Based MARL & Learn a world model from data; agents learn from the learned model to avoid direct interaction with the environment, improving sample efficiency & MAMBPO\cite{willemsen2021mambpo}, AORPO~\cite{DBLP:conf/ijcai/0001WSZ21}, MBVD~\cite{xu2022mingling}, MAMBA~\cite{egorov2022scalable}, VDFD~\cite{wang2023leveraging} & Significant improvement in sample utilization efficiency and cooperative effectiveness in complex scenarios with the help of successful model learning methods or the development of methods tailored for MASs \\
    \bottomrule
    \end{tabularx}}
    \label{closeenvironment}
\end{table*}

\begin{table*}
\renewcommand{\arraystretch}{1}
\centering
\caption{Research directions in cooperative multi-agent reinforcement learning in classical environments (continued).}
\resizebox{0.94\textwidth}{!}{
\begin{tabularx}{\textwidth}{YZYZ}
\toprule
Research Direction & Core Content & Representative Algorithms & Applications and Achievements \\
\midrule
Action Hierarchical Learning & Decompose complex problems into multiple sub-problems, solve each sub-problem separately, and then solve the original complex problem & FHM~\cite{ahilan2019feudal}, HSD~\cite{yang2020hierarchical}, RODE~\cite{wang2020rode}, ALMA~\cite{iqbal2022alma}, HAVEN~\cite{xu2023haven}, ODIS~\cite{zhang2023discovering} & Significant improvement in the coordination efficiency of MASs in various task scenarios \\
\midrule
Topological Structure Learning & Model interaction relationships between multiple agents, using coordination graphs, and other means to describe the interaction relationships between agents & CG~\cite{DBLP:conf/icml/GuestrinLP02}, DCG~\cite{dcg}, DICG~\cite{li2021deep}, MAGIC~\cite{niu2021multi}, ATOC~\cite{DBLP:conf/nips/JiangL18}, CASEC~\cite{DBLP:conf/iclr/00010DY0Z22} & Implicit or explicit representation of the relationships between agents; can reduce the joint action space in complex scenarios, improving cooperative performance \\
\midrule
Other Aspects & Research on interpretability, theoretical analysis, social dilemmas, large-scale scenarios, delayed rewards, etc. & Na2q~\cite{liu2023n}, ACE~\cite{li2023ace}, CM3~\cite{yang2019cm3}, MAHHQN~\cite{fu2019deep}, references~\cite{chen2021variational, da2020agents, grupen2022cooperative, da2017moo} & Further comprehensive research on cooperative MARL \\
\bottomrule
\end{tabularx}}
\label{closeenvironmentcon}
\end{table*}
\subsubsection{Cooperative Multi-Agent Exploration}

RL methods strive to efficiently learn optimal policies, where high-quality training samples are crucial, and exploration plays a key role during the sampling process, constituting a vital aspect of RL~\cite{hao2023exploration}. Similar to exploration techniques widely used in single-agent scenarios, exploration in cooperative MARL has garnered attention. Initial algorithms like QMIX and MADDPG, designed for multi-agent tasks, lacked specific exploration policies and underperformed in certain complex scenarios. Subsequent works aim at addressing multi-agent exploration challenges. MAVEN~\cite{DBLP:conf/nips/MahajanRSW19} introduced a latent variable to maximize mutual information between this variable and generated trajectories, addressing the exploration limitations of value decomposition methods like QMIX. EITI and EDTI~\cite{wang2019influence} proposed cooperative exploration methods by considering interactions among agents, demonstrating improved coordination in various environments. CMAE~\cite{liu2021cooperative} uses constrained space selection to encourage exploration in regions with higher value, enhancing cooperative efficiency. EMC~\cite{zheng2021episodic} extends single-agent, motion-based exploration methods to the multi-agent domain and proposes episodic-memory-based approaches to store high-value trajectories for exploration facilitation. Uneven~\cite{gupta2021uneven} improves exploration efficiency by simultaneously learning multiple task sets, obtaining a multi-agent generalization feature to address relative overgeneralization in multi-agent scenarios. SMMAE~\cite{zhang2023self} balances individual and team exploration to enhance coordination efficiency, achieving satisfactory results in complex tasks. Recently developed ADER~\cite{DBLP:conf/icml/KimS23} focused on studying the balance between exploration and exploitation in cooperative multi-agent tasks.

Additionally, various methods have explored multi-agent exploration from different perspectives, such as intrinsic reward exploration~\cite{iqbal2019coordinated}, decentralized multi-agent exploration in distributed scenarios~\cite{viseras2016decentralized,he2023decentralized}, information-based exploration methods~\cite{baglietto2002information}, cooperative exploration under unknown initial conditions~\cite{yan2023mui}, exploration problems in multi-agent multi-armed bandit settings~\cite{chakraborty2017coordinated}, exploration-exploitation balance in competitive multi-agent scenarios~\cite{DBLP:conf/nips/LeonardosPS21}, structure-based exploration methods~\cite{jo2023fox}, optimistic exploration~\cite{oh2023toward}, and exploration in reward-sparse scenarios~\cite{xuexploration}. While these methods have shown promising results in diverse testing environments, developing test environments tailored for various exploration algorithms and proposing comprehensive theories for multi-agent exploration remain worthwhile areas for future research.

\subsubsection{Multi-Agent Communication}

\begin{figure*}
\centering
\includegraphics[width=0.8\linewidth]{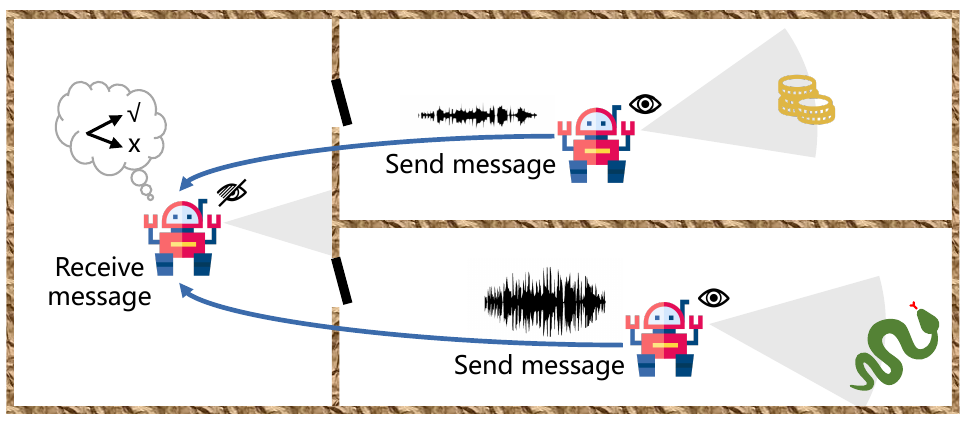}
\caption{Cooperative Multi-Agent Communication.}\label{communication}
\end{figure*}

Communication plays a crucial role in MASs, enabling agents to share experiences, intentions, observations, and other vital information based on communication technology (Figure~\ref{communication}~\cite{zhu2022survey}). Three factors typically characterize multi-agent communication:  when, with which  agents, and what type of information is exchanged. Early research focused on integrating communication with existing MARL to improve coordination efficiency. DIAL~\cite{foerster2016learning} introduced a simple communication mechanism where agents broadcast information to all teammates, facilitating end-to-end RL training. CommNet~\cite{communication16} proposed an efficient centralized communication structure where hidden layer outputs of all agents are collected and averaged to enhance local observations. VBC~\cite{vbc} adds a variance-based regularizer to eliminate noise components in information, achieving lower communication overhead and better performance than other methods. IC3Net~\cite{ic3net}, Gated-ACML~\cite{acml}, and I2C~\cite{i2c} learn gating mechanisms to decide when and with which teammates to communicate, reducing information redundancy. SMS~\cite{di2022yang} models multi-agent communication as a cooperative game, evaluating the importance of each message for decision-making and trimming channels with no positive gain. MASIA~\cite{guan2023efficient} introduced the concept of information aggregation, extracting information effectively for decision-making at the information receiver. Additionally, it designed an open-source offline communication dataset, demonstrating the efficiency and rationality of the proposed communication structure in various experiments.

Some works have focused on determining the content of communication. TarMAC~\cite{tarmac} generate information by appending signatures, and the receiver use attention mechanisms to calculate the similarity between its local observation and the information signatures from different teammates. DAACMP~\cite{doubleattention}, based on the MADDPG~\cite{maddpg} framework, introduced a dual attention mechanism, showing that attention improves the coordination performance of MASs. NDQ~\cite{ndq} uses information-theoretic regularization to design two different information-theoretic regularization terms for optimizing the value function, achieving good coordination performance in locally observable scenarios. TMC~\cite{tmc} applied smoothing and action selection regularizers for concise and robust communication, adding received information as an incentive to individual value functions. MAIC~\cite{maic} designed a decentralized incentive communication based on teammate modeling, demonstrating excellent communication efficiency in various task scenarios. Recent works, including CroMAC~\cite{yuan2023comrobust}, AME\cite{sun2022certifiably}, $\mathcal{R}$-MACRL~\cite{xue2022mis}, and MA3C~\cite{yuan2023communication} focus on communication policy robustness during deployment, addressing potential communication interference issues. While existing methods have demonstrated effectiveness in various task scenarios, the natural gap between multi-agent environments and human society poses challenges. Exploring ways to bridge this gap through technologies like large language models~\cite{pang2023language}, promoting human-AI interaction, and conducting interpretable research on communication content are promising areas for future study.

\subsubsection{Cooperative Multi-Agent Reinforcement Learning with Agent Modeling}

\begin{figure*}
\centering
\includegraphics[width=0.8\linewidth]{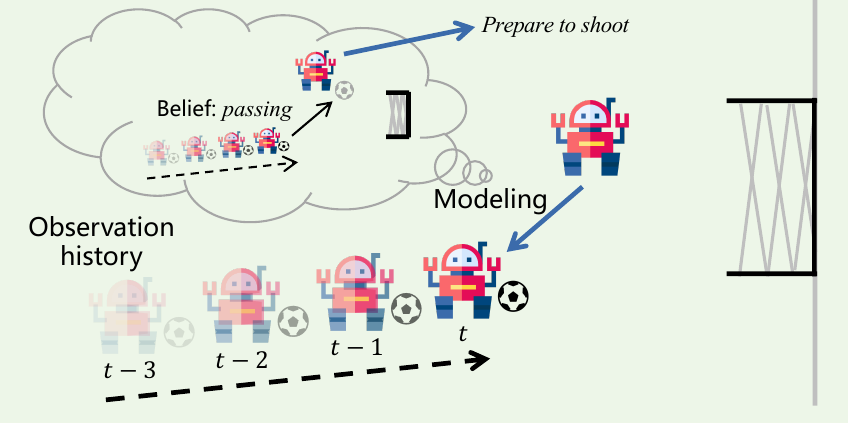}
\caption{MARL with Agent Modeling.}\label{agentmodel}
\end{figure*}

Endowing agents with the ability to infer actions, goals, and beliefs of other agents in the environment is a key focus in MARL, particularly in cooperative tasks~\cite{albrecht2018autonomous}. Effective and accurate modeling techniques enable agents to collaborate efficiently (Figure~\ref{agentmodel}). One direct approach is employing Theory of Mind (ToM) from psychology, where ToMnet~\cite{rabinowitz2018machine}, and subsequent works like ToM2C~\cite{wang2021tom2c}, CTH~\cite{shum2019theory}, and \cite{DBLP:conf/hhai/ErdoganDVY22}, incorporate psychological theories into multi-agent tasks, enhancing teammate modeling and improving coordination capabilities. More details and development of ToM can be seen in~\cite{langley2022theory}.

Alternatively, some works employ different techniques to model other teammates. OMDDPG~\cite{papoudakis2020variational} uses Variational Autoencoders (VAEs) to infer the behavior of other agents based on local information. LIAM~\cite{papoudakis2021agent} extended this technique to locally observable conditions, developing an efficient teammate modeling technique. LINDA~\cite{linda} mitigates local observability issues by considering perceived teammate information, assisting in enhancing MAS coordination across diverse tasks. MACC~\cite{macc} uses local information learning for subtask representation to augment individual policies. MAIC~\cite{maic} introduced a communication-based method for teammate modeling, extracting crucial information for efficient decision-making. LILI~\cite{xie2021learning} considered agent modeling in robotic scenarios and developed algorithms based on high-level policy representation learning to promote coordination. SILI~\cite{wang2022influencing} further considered learning a smooth representation to model other agents based on LILI. Literature~\cite{tian2023towards} empowered robots to understand their own behaviors and develop technologies to better assist humans in achieving rapid value alignment.

In addition to the modeling techniques mentioned above, SOM~\cite{raileanu2018modeling} enables agents to learn how to predict the behaviors of teammates by using their own policies, and update their unknown target beliefs about other teammates online accordingly. MBOM~\cite{yu2022model} developed adversary modeling techniques based on model learning. \cite{hernandez2019agent} takes modeling the behavior of other agents as an additional objective when optimizing the policy. DPIQN and DRPIQN~\cite{hong2018deep} considered modeling multi-agent task scenarios with policy changes and propose using a policy feature optimizer to optimize policy learning. ATT-MADDPG~\cite{mao2019modelling} obtains a joint teammate representation to evaluate teammate policies and promote multi-agent coordination via attention based MADDPG~\cite{DBLP:journals/corr/LillicrapHPHETS15}. Literature~\cite{wen2018probabilistic} and subsequent work considered solving the cyclic inference problem in teammate modeling through techniques such as probability inference. Literature~\cite{roy2020promoting} proposed TeamReg and CoachReg to evaluate the action choices of the team and achieve better coordination. The aforementioned works have shown good coordination performance in various task scenarios. However, in these works, agents need to model other agents or entities in the environment. When dealing with large-scale agent systems or a large number of entities, using above methods will greatly increase computational complexity. Developing technologies such as distributed grouping to improve computational efficiency is one of the future research directions and valuable topics.

\subsubsection{Multi-Agent Policy Imitation}

Imitation Learning (IL)~\cite{zare2023survey} offers a method for agents to make intelligent decisions by mimicking human experts. In multi-agent scenarios, IL has advanced significantly. For example, \cite{le2017coordinated} proposed to learn a probabilistic cooperative model in latent space from demonstration data. \cite{zhan2018generating} extended this by using a hierarchical structure for high-level intention modeling. \cite{wang2021multi1} employed a connection structure~\cite{nelsen2006introduction} to explicitly model coordination relationships. Multi-agent Inverse RL (IRL) has also been explored, with MAGAIL~\cite{DBLP:conf/nips/SongRSE18} extending adversarial generative techniques to MASs. MA-AIRL~\cite{yu2019multi} developed a scalable algorithm for high-dimensional state spaces and unknown dynamic environments. \cite{gruver2020multi} learns a reward function based on the representation space to represent the behavior of agents. CoDAIL~\cite{liu2019multi} introduced a distributed adversarial imitation learning method based on correlated policies. DM$^2$~\cite{wang2022dm} recently presented a distributed multi-agent imitation learning method based on distribution matching. MIFQ~\cite{bui2023inverse} studied a decomposed multi-agent IRL method, demonstrating excellent performance in various scenarios.

Other research have explored multi-agent imitation learning from different perspectives, including applications in imitatiing driving~\cite{bhattacharyya2018multi}, policy research in multi-agent IRL~\cite{fu2021evaluating}, multi-agent IRL without demonstration data~\cite{wang2018competitive}, application to mean field game problems~\cite{chen2023adversarial}, design for mulit-agent systems in distributed constrained IRL~\cite{liu2022distributed}, and asynchronous multi-agent imitation learning~\cite{zhang2021imitation}. Despite progress in various aspects, there's a need for research exploring large-scale scenarios, handling heterogeneous or suboptimal demonstration data, and applying multi-agent imitation learning to real-world scenarios like autonomous driving. Investigating these aspects and fostering coordination between AI and domain experts can drive significant advancements in multi-agent policy imitation.

\subsubsection{Model-Based Cooperative MARL}
Reinforcement learning, due to its need for continual interaction with the environment for policy optimization, often employs model-free algorithms, where agents use samples from interaction with the environment. However, these algorithms frequently suffer from low sample efficiency. Model-based RL, considered to have higher sample efficiency, typically involves learning the environment's state transition function (often including the reward function), i.e., the world model. Subsequently, policy optimization is conducted based on the knowledge derived from the world model~\cite{luo2022survey,moerland2023model}, leading to a significant improvement in sample efficiency.

In recent years, model-based MARL has garnered attention and made progress~\cite{wang2022model}. Early efforts extended successful techniques in single-agent scenarios to multi-agent settings. For instance, MAMBPO~\cite{willemsen2021mambpo} extended the MBPO method~\cite{DBLP:conf/nips/JannerFZL19} to multi-agent tasks, constructing a model-based RL system based on the CTDE paradigm and a world model. This approach moderately enhances the system's sampling efficiency. CTRL~\cite{park2019multi} and AORPO~\cite{DBLP:conf/ijcai/0001WSZ21} further introduced opponent modeling techniques into model learning to acquire adaptive trajectory data, augmenting the original data. MARCO~\cite{zhang2022centralized} learns a multi-agent world model and employs it to learn a centralized exploration policy to collect more data in high uncertainty areas, enhancing multi-agent policy learning. Some techniques extend widely used single-agent methods, such as Dreamer~\cite{hafner2019dream}, to multi-agent tasks. MBVD~\cite{xu2022mingling} learns an implicit world model based on value decomposition, enabling agents to evaluate state values in latent space, providing them with foresight. MAMBA~\cite{egorov2022scalable} enhances centralized training in cooperative scenarios using model-based RL. The study discovers that algorithm optimization can be achieved during training through a world model and, during execution, a new corresponding world model can be established through communication. VDFD~\cite{wang2023leveraging} significantly improves the sample efficiency of MARL by developing techniques that decouple the learning of the world model.

Additionally, researchers explore the learning of world models in MASs from various perspectives, such as collision avoidance in distributed robot control~\cite{wang2021model}, safety and efficiency improvement~\cite{xiao2023model}, agent interaction modeling~\cite{mahajan2021tesseract}, optimization of distributed network systems~\cite{DBLP:conf/iros/DuMLLDW022}, model-based opponent modeling~\cite{yu2022model}, efficiency improvement in multi-agent communication through model-based learning~\cite{kim2020communication,ding2022multi,han2023model}, model-based mean-field MARL~\cite{pasztor2023efficient}, efficient multi-agent model learning~\cite{sessa2022efficient}, improving the learning efficiency of offline MARL through model learning~\cite{barde2023model}, and applications of model-based MARL~\cite{han2022multiagent}. Although some of these methods have achieved certain results, the challenges posed by the curse of dimensionality with the increasing number of agents and the partial observability introduced by decentralized execution often hinder the development of model-based MARL algorithms. Exploring efficient methods tailored to the characteristics of MASs, such as subset selection~\cite{DBLP:conf/nips/QianYZ15}, is an essential research direction.

\subsubsection{Multi-Agent Hierarchical Reinforcement Learning and Skill Learning}
Reinforcement learning faces the challenge of the curse of dimensionality in complex scenarios. To address this issue, researchers have proposed hierarchical RL (HRL)~\cite{pateria2021hierarchical}. The primary goal of HRL is to decompose complex problems into smaller ones, solve each subproblem independently, and thus achieve the overall goal. In recent years, there have been advancements in multi-agent hierarchical RL~\cite{makar2001hierarchical}. FHM~\cite{ahilan2019feudal} introduced a feudal lord structure into the process of MARL. This method is designed for scenarios where agents have different task objectives but may not be suitable for cooperative tasks with shared objectives. To address the sparse and delayed reward issues in multi-agent tasks, \cite{tang2018hierarchical} proposed a hierarchical version of the QMIX algorithm and a hierarchical communication structure. This method achieves excellent cooperation performance in various task scenarios but requires manual design of high-level actions. HSD~\cite{yang2020hierarchical} generates corresponding skills through an high-level macro policy, and this method is mainly trained through supervised learning. RODE~\cite{wang2020rode} learns action representations and selects actions hierarchically to enhance cooperation performance in various scenarios. VAST~\cite{phan2021vast} addresses the efficiency issues in large-scale multi-agent cooperation tasks through a hierarchical approach. ALMA~\cite{iqbal2022alma} fully exploits the task structure in multi-agent tasks and uses high-level subtask decomposition policies and lower-level agent execution policies. HAVEN~\cite{xu2023haven} introduces a bidirectional hierarchical structure for both inter-agent and intra-agent coordination to further improve multi-agent cooperation efficiency.

Apart from the mentioned works, some skill-based approaches in SARL have made progress~\cite{eysenbach2018diversity} and have been applied to MARL. \cite{friedrich1998integrating} considered integrating skills into MASs to improve overall performance. HSL~\cite{liu2022heterogeneous} designs a method to extract heterogeneous and sufficiently diverse skills from various tasks for downstream task learning. ODIS~\cite{zhang2023discovering} focuses on extracting skills from offline datasets. MASD~\cite{he2020skill} aims to discover skills beneficial for cooperation by maximizing the mutual information between the potential skill distribution and the combination of all agent states and skills. SPC~\cite{wang2023towards} focuses on the automatic curriculum learning process of skill populations in MARL. \cite{chen2022scalable} also explores learning to augment multi-agent skills. While these methods to some extent learn or use skills in MASs, the interpretability of the learned skills is currently lacking. Exploring ways to make the learned skills interpretable, such as bridging the gap between MASs and the human society through natural language~\cite{pang2023natural}, is a topic worth researching.

\subsubsection{Cooperative Multi-Agent Reinforcement Learning Topology Structure Learning}
The interaction among agents is a focal point in the study of multi-agent problems. A coordination graph is a method that explicitly characterizes the interaction between agents by decomposing the multi-agent value function into a graph representation~\cite{DBLP:conf/aaai/GuestrinVK02,DBLP:conf/icml/GuestrinLP02}. In recent years, this approach has been widely applied in cooperative MARL. In coordination graph-based MARL tasks, nodes generally represent agents, and edges (hyper-edges) represent connected relationships between agents, constructing a reward function on the joint observation action space. A coordination graph can represent a high-order form of value decomposition among multiple agents. Typically, Distributed Constraint Optimization (DCOP) algorithms~\cite{cheng2012coordinating} are employed to find action selections that maximize the value. Agents can exchange information through connected edges over multiple rounds. DCG~\cite{dcg} introduces some deep learning techniques into coordination graphs, expanding them into high-dimensional state-action spaces. In complex task scenarios, such as SMAC, DCG can relieve the relative overgeneralization issues among multiple agents. However, DCG mainly focuses on pre-training on static and dense topological structures, exhibiting poor scalability and requiring dense and inefficient communication methods in dynamic environments.

A core issue in coordination graph problems is how to learn a dynamic and sparse graph structure that satisfies agent action selection. Sparse cooperative Q-function learning~\cite{kok2006collaborative} attempts to learn sparse graph structures for value function learning. However, the graph structure learned by this method is static and requires substantial prior knowledge. Literature~\cite{zhang2013coordinating} proposed learning a minimal dynamic graph set for each agent, but the computational complexity exponentially increases with the number of agent neighbors. Literature~\cite{castellini2019representational} studies the expressiveness of some graph structures but mainly focuses on random topologies and stateless problems. Literature~\cite{DBLP:conf/iclr/00010DY0Z22} proposed a new coordination graph testing environment, MACO, and introduced the Context-Aware Sparse Deep Coordination Graphs algorithm (CASEC). This algorithm effectively reduces the reward function evaluation error introduced during the graph construction process by learning induced representations and performs well in multiple environments. Subsequent work also enhances coordination from multiple perspectives, such as developing nonlinear coordination graph structures~\cite{kang2022non}, self-organizing polynomial coordination graph structures~\cite{yang2022self}, and so on.

On the other hand, some methods use techniques such as attention mechanisms to find implicit graph structures among multiple agents by cutting unnecessary connection structures through attention mechanisms. For instance, ATOC~\cite{DBLP:conf/nips/JiangL18} learns the most essential communication structure through attention mechanisms. DICG~\cite{li2021deep} learns implicit coordination graph structures through attention mechanisms. MAGIC~\cite{niu2021multi} improves the communication efficiency and teamwork of MASs through graph attention mechanisms. Although these methods can to some extent obtain or improve the interaction topology between agents, currently, these methods generally make progress only in scenarios with a small number of agents. The challenge of obtaining the optimal topology structure in large-scale and strongly interacting scenarios is a future research direction~\cite{DBLP:conf/atal/ShengW0LWYCZ23}.

\subsubsection{Other Aspects}
In addition to the extensively researched topics mentioned above, several other aspects are gradually being explored and uncovered in MARL. These include but are not limited to cooperative interpretability~\cite{liu2023n}, developing multi-agent cooperation algorithms with decision-making order~\cite{kuba2021settling,wen2022multiagent,li2023ace,wang2023order}, theoretical analysis of multi-agent cooperation~\cite{wang2021towards,dou2022understanding,hu2003multi}, multi-agent multi-task (stage) cooperation learning~\cite{yang2019cm3}, social dilemma problems in MASs~\cite{leibo2017multi}, asynchronous multi-agent cooperation~\cite{xiao2022asynchronous,zhang2023himacmic}, large-scale multi-agent cooperation~\cite{fu2022concentration}, delayed reward in multi-agent cooperation~\cite{qiu2022off}, multi-agent cooperation in mixed action spaces~\cite{fu2019deep}, discovery of causal relationships in multi-agent cooperation~\cite{grimbly2021causal}, curriculum learning in multi-agent cooperation~\cite{chen2021variational}, research on teacher-student relationships in cooperative MARL~\cite{da2020agents}, fairness in MASs~\cite{grupen2022cooperative}, and entity-based multi-agent cooperation~\cite{da2017moo}, among others.

\subsection{Typical Benchmark Scenarios}
While designing and researching algorithms, some works have developed a series of test environments to comprehensively evaluate algorithms from various aspects. Typical environments include the StarCraft Multi-Agent Challenge (SMAC)~\cite{pymarl} and its improved version SMACv2~\cite{ellis2022smacv2}, SMAClite~\cite{michalski2023smaclite}, Multi-Agent Particle World (MPE)~\cite{maddpg}, Multi-Agent MuJoCo (MAMuJoCo)~\cite{peng2021facmac}, Google Research Football (GRF)~\cite{kurach2020google}, Large-scale Multi-Agent Environment (MAgent)~\cite{zheng2018magent}, Offline MARL Test Environment~\cite{formanek2023off}, and Multi-Agent Taxi Environment TaxAI~\cite{mi2023taxai}, among others. Common multi-agent test environments and their characteristics are shown in Table~\ref{table:benchmark}. Moreover, for the convenience of future work, some researchers have integrated and open-sourced the current mainstream testing environments, including Pymarl~\cite{pymarl} \footnote{https://github.com/oxwhirl/pymarl}, EPyMARL~\cite{lbf}\footnote{https://github.com/uoe-agents/epymarl}, PettingZoo~\cite{terry2021pettingzoo}\footnote{https://github.com/Farama-Foundation/PettingZoo}, MARLlib~\cite{hu2022marllib}\footnote{https://marllib.readthedocs.io/en/latest/index.html}, and the environment from the 1st AI Agents Contest\footnote{http://www.jidiai.cn/environment}, among others.

\begin{table*}
    \centering
    \caption{Introduction to Typical Multi-Agent Testing Environments.}
    \resizebox{\textwidth}{!}{
    \begin{tabular}{cccccccc}
    \toprule
    Environment & \begin{tabular}[x]{@{}c@{}}Heterogeneous\\Agents\end{tabular} & \begin{tabular}[x]{@{}c@{}}Scenario\\Type\end{tabular} & \begin{tabular}[x]{@{}c@{}}Observation\\Space\end{tabular} & \begin{tabular}[x]{@{}c@{}}Action\\Space\end{tabular} & \begin{tabular}[x]{@{}c@{}}Typical\\Number\end{tabular} & \begin{tabular}[x]{@{}c@{}}Communication\\Capability\end{tabular} & Problem Domain\\
    \midrule
        \makecell{Matrix Games~\cite{claus1998dynamics}\\ (1998)}
         & Yes & Mixed & Discrete & Discrete & 2 & No & Matrix Games \\
        \midrule
        \makecell{MPE~\cite{maddpg}\\ (2017)} & Yes & Mixed & Continuous & Discrete & 2-6 & Allowed & Particle Games \\
        \makecell{MACO~\cite{DBLP:conf/iclr/00010DY0Z22}\\ (2022)}  & No & Mixed & Discrete & Discrete & 5-15 & Allowed & Particle Games \\
        \makecell{GoBigger~\cite{zhang2022gobigger}\\ (2022)}  & No & Mixed & Continuous & Continuous or Discrete & 4-24 & No & Particle Games\\
        \makecell{MAgent~\cite{zheng2018magent}\\ (2018)} & Yes & Mixed & Continuous+Image & Discrete & 1000 & No & \makecell{Large-Scale \\Particle Confrontation} \\
        \midrule
        \makecell{MARLÖ ~\cite{perez2019multi} \\(2018)} & No & Mixed & Continuous+Image & Discrete & 2-8 & No & Adversarial Games \\
        \makecell{DCA \cite{fu2022concentration} \\(2022)}  & No & Mixed & Continuous & Discrete & 100-300 & No & Adversarial Games \\
        \makecell{Pommerman~\cite{resnick2018pommerman} \\(2018)} & No & Mixed & Discrete & Discrete & 4 & Yes & Bomberman Game \\
        \makecell{SMAC ~\cite{pymarl} \\(2019)} & Yes & Cooperative & Continuous & Discrete & 2-27 & No & StarCraft Game \\
        \makecell{Hanabi ~\cite{bard2020hanabi} \\(2019)} & No & Cooperative & Discrete & Discrete & 2-5 & Yes & Card Game \\
        \makecell{Overcooked \cite{DBLP:conf/nips/CarrollSHGSAD19} \\(2019)}  & Yes & Cooperative & Discrete & Discrete & 2 & No & Cooking Game \\
        \makecell{Neural MMO ~\cite{suarez2021neural} \\(2019)} & No & Mixed & Continuous & Discrete & 1-1024 & No & Multiplayer Game \\
        \makecell{Hide-and-Seek~\cite{baker2019emergent} \\(2019)} &  Yes  & Mixed & Continuous &  Discrete &   2-6 & No & Hide and Seek Game  \\
        \makecell{LBF ~\cite{lbf} \\(2020)} & No & Cooperative & Discrete & Discrete & 2-4 & No & Food Search Game \\
        \makecell{Hallway~\cite{ndq} \\(2020)}  & No & Cooperative & Discrete & Discrete & 2 & Yes & \makecell{Communication \\Corridor Game} \\
        \midrule
        \makecell{GRF ~\cite{kurach2020google} \\(2019)} & No & Cooperative & Continuous & Discrete & 1-3 & No & Soccer Confrontation \\
        \makecell{Fever Basketball~\cite{jia2020fever} \\(2020)} &  Yes  & Mixed  & Continuous & Discrete & 2-6  & No & Basketball Confrontation  \\
        \midrule
        \makecell{SUMO~\cite{krajzewicz2010traffic} \\(2010)}  & No & Mixed & Continuous & Discrete & 2-6 & No & Traffic Control \\
        \makecell{Traffic Junction\cite{communication16} \\(2016)}  & No & Cooperative & Discrete & Discrete & 2-10 & Yes & \makecell{Communication \\Traffic Scheduling} \\
        \makecell{CityFlow~\cite{DBLP:conf/www/ZhangFLDZZ00JL19} \\(2019)}  & No & Cooperative & Continuous & Discrete & 1-50+  & No & Traffic Control \\
        \makecell{MAPF \cite{DBLP:conf/socs/SternSFK0WLA0KB19} \\(2019)}  & Yes & Cooperative & Discrete & Discrete & 2-118 & No & Path Navigation \\
        \makecell{Flatland ~\cite{mohanty2020flatland} \\(2020)} & No & Cooperative & Continuous & Discrete & $>$100 & No & Train Scheduling \\
        \makecell{SMARTS ~\cite{zhou2020smarts} \\(2020)} & Yes & Mixed & Continuous+Image & Continuous or Discrete & 3-5 & No & Autonomous Driving \\
        \makecell{MetaDrive \cite{li2022metadrive} \\(2021)}  & No & Mixed & Continuous & Continuous & 20-40 & No & Autonomous Driving \\
        \makecell{MATE \cite{pan2022mate} \\(2022)} & Yes & Mixed & Continuous & Continuous or Discrete & 2-100+ & Yes & Target Tracking \\
        \makecell{MARBLER~\cite{torbati2023marbler} \\(2023)}  & Yes & Mixed & Continuous & Discrete & 4-6 & Allowed & Traffic Control \\
        \midrule
        \makecell{RWARE~\cite{lbf} \\(2020)} & No & Cooperative & Discrete & Discrete & 2-4 & No & Warehouse Logistics \\
        \makecell{MABIM~\cite{yang2023versatile}\\ (2023)}  & No  & Mixed & Continuous & Continuous or Discrete & 500-2000 & No & Inventory Management  \\
        \makecell{MaMo \cite{xue2022multi} \\(2022)}  & Yes & Cooperative & Continuous & Continuous & 2-4 & No & Parameter Tuning \\
        \makecell{Active Voltage Control \cite{wang2021multi}\\ (2021)}  & Yes & Cooperative & Continuous & Continuous & 6-38 & No & Power Control \\
        MAMuJoCo ~\cite{peng2021facmac} (2020) & Yes & Cooperative & Continuous & Continuous & 2-6 & No & Robot Control \\
        \midrule
        \makecell{Light Aircraft Game~\cite{liu2022light}\\(2022)}  & No & Mixed & Continuous & Discrete & 1-2 & No & Intelligent Air Combat \\
        \makecell{MaCa~\cite{gao2019maca} \\(2020)}  & Yes & Mixed & Image & Discrete & 2 & No & Intelligent Air Combat \\
        \midrule
        \makecell{Gathering~\cite{leibo2017multi} \\(2020)}  & No & Cooperative & Image & Discrete & 2 & No & Social Dilemma \\
        \makecell{Harvest~\cite{harvest} \\(2017)}  & No & Mixed & Image & Discrete & 3-6 & No & Social Dilemma \\
        \midrule
        \makecell{Safe MAMuJoCo~\cite{gu2023safe} \\(2023)} & Yes  & Cooperative & Continuous & Continuous & 2-8 & No & Safe Multi-Agent  \\ 
        \makecell{Safe MARobosuite~\cite{gu2023safe}\\(2023)} & Yes  & Cooperative & Continuous & Continuous & 2-8 & No & Safe Multi-Agent  \\ 
        \makecell{Safe MAIG~\cite{gu2023safe} \\(2023)} & Yes  & Cooperative & Continuous & Continuous & 2-12 & No & Safe Multi-Agent  \\  
        \midrule
        \makecell{OG-MARL~\cite{formanek2023off} \\(2023)} & Yes  & Mixed & Continuous & Continuous or Discrete & 2-27 & No & Offline Dataset  \\ 
        \makecell{MASIA~\cite{guan2023efficient} \\(2023)} & Yes  & Cooperative & Discrete or Continuous & Discrete & 2-11 & No & Offline Communication Dataset  \\ 
        \bottomrule
    \end{tabular}}
    \label{table:benchmark}
\end{table*}

\subsection{Typical Application Scenarios}
Meanwhile, cooperative MARL algorithms have been widely applied in various task scenarios, including gaming, industrial applications, robot control, cross-disciplinary applications, and military domains~\cite{bahrpeyma2022review,canese2021multi,oroojlooy2023review,li2022applications,zhou2023multi}.

Early multi-agent algorithms primarily focused on the gaming domain, optimizing policies using RL algorithms. AlphaStar~\cite{vinyals2019grandmaster}, based on RL and employing multi-agent learning algorithms like population training, demonstrated remarkable performance in controlling in-game agents to defeat opponents in StarCraft~\cite{yang2020overview,li2023combining}. Subsequent works extended MARL algorithms to other gaming tasks. The ViVO team achieved effective adversarial results in controlling hero units in the game Honor of Kings using hierarchical RL in various scenarios~\cite{zhang2019hierarchical}. Many test environments, including SMAC~\cite{pymarl}, are developed based on game engines. Besides real-time games, MARL has also shown success in non-real-time games such as chess~\cite{perolat2022mastering}, Chinese chess~\cite{li2023jiangjun}, mahjong~\cite{zhao2022towards}, poker~\cite{zha2021douzero}, football~\cite{kurach2020google}, basketball games~\cite{jia2020fever,yeh2019diverse}, and hide-and-seek~\cite{baker2019emergent}.

On the other hand, researchers have explored applying MARL to industrial domains. Leveraging the significant potential of MARL in problem-solving, studies model industrial problems as MARL tasks. For example, in~\cite{troullinos2021collaborative}, unmanned driving is modeled as a cooperative problem, promoting cooperation between unmanned vehicles using a coordination graph. The study found that automatic vehicle scheduling can be achieved in various task scenarios. Other related works use MARL to enhance traffic signal control~\cite{wang2021adaptive}, unmanned driving~\cite{chen2022multi}, drone control~\cite{jeon2022multiagent,chen2023robust}, and autonomous vehicle control~\cite{choi2022marl}. Additionally, researchers have explored power-related applications, such as controlling the frequency of wind power generation using MARL algorithms~\cite{liang2022multiagent}. Some works have delved into finance, applying MARL in various scenarios~\cite{huang2022correction,shavandi2022multi,huang2023multi}. Furthermore, MAAB~\cite{wen2022cooperative} designs an online automatic bidding framework based on MARL, addressing other issues~\cite{fang2023learning}. In addition, some works synchronize clocks on FPGAs using MARL~\cite{cardarilli2022fpga}, and in the context of virtual Taobao, MARL is applied for better capturing user preferences~\cite{shi2019virtual}. Some studies focus on controlling robots using MARL~\cite{ismail2018survey,rasheed2022review,dahiya2023survey}, such as~\cite{scheikl2021cooperative}, applying cooperative MARL to cooperative assistance tasks in robot surgery, significantly improving task completion.

MARL is also applied in cross-disciplinary fields. For instance, MA-DAC~\cite{xue2022multi} models multi-parameter optimization problems as MARL tasks and solves them using cooperative MARL algorithms like QMIX, significantly enhancing the ability to optimize multiple parameters. MA2ML~\cite{wang2023multi} effectively addresses optimization learning problems in the connection of modules in automated machines using MARL. The study in~\cite{wang2021collaborative} enriches MARL under rich visual information and designs navigation tasks in a cooperative setting. Literature~\cite{fang2022coordinate} proposed a multi-camera cooperative system based on cooperative alignment to solve the active multi-target tracking problem. Literature~\cite{lin2023local} modeled image data augmentation as a multi-agent problem, introducing a more fine-grained automatic data method by dividing an image into multiple grids and finding a jointly optimal enhancement policy. Additionally, many works focus on applying MARL algorithms to combinatorial optimization. Literature~\cite{zhang2022online} studies optimizing online parking allocation through MARL, literature~\cite{sui2020multi} focuses on lithium-ion battery scheduling using a MARL framework, literature~\cite{wang2022solving} leverages MARL to solve job scheduling problems in resource preemptive environments, and literature~\cite{zhou2021multi} explores online scheduling problems in small-scale factories using MARL. MARLYC~\cite{kadoche2023marlyc} proposed a new method called MARL yaw control to suggest controlling the yaw of each turbine, ultimately increasing the total power generation of the farm. Cooperative MARL has also found applications in daily life, such as~\cite{ijcai2023p38}, testing the acoustic effects in a room using two collaborating robots, and~\cite{xu2020multi}, managing energy in daily home environments. Literature~\cite{zhou2018botzone} explores online intelligent education based on MARL. Some studies also attempt to apply multi-agent collaboration techniques to the medical field, such as using multi-agent collaboration for neuron segmentation~\cite{chenself} or medical image segmentation~\cite{allioui2022multi}.

In addition to the applications mentioned earlier, MARL has also been explored in the field of national defense applications~\cite{gong2023uav,li2023multi,liu2021graph,basak2022utility,weiren2020air,piao2023spatio,li2023multi}. The work in~\cite{jiang2021anti} proposes an approach based on MADDPG and attention mechanisms capable of handling scenarios with variable teammates and opponents in multi-agent aerial combat tasks. In~\cite{yue2022unmanned}, a hierarchical MARL (HMARL) approach is introduced to address the cooperative decision-making problem of heterogeneous drone swarms, specifically focusing on the Suppression of Enemy Air Defenses (SEAD) mission by decoupling it into two sub-problems: high-level target assignment (TA) and low-level cooperative attack (CA).
The study in~\cite{jiandong2021uav} constructs a multi-drone cooperative aerial combat system based on MARL. Simulation results indicate that the proposed strategy learning method can achieve a significant energy advantage and effectively defeat various opponents.
In~\cite{kong2023hierarchical}, a hierarchical MARL framework for air-to-air combat is proposed to handle scenarios involving multiple heterogeneous intelligent agents. MAHPG~\cite{sun2021multi} designs a strategy gradient MARL method based on self-play adversarial training and hierarchical decision networks to enhance the aerial combat performance of the system, aiming to learn various strategies. Additionally, using hierarchical decision networks to handle complex mixed actions, the study in~\cite{sun2023multi} designs a mechanism based on a two-stage graph attention neural network to capture crucial interaction relationships between intelligent agents in intelligent aerial combat scenarios. Experiments demonstrate that this method significantly enhances the system's collaborative capabilities in large-scale aerial combat scenarios.

\section{Cooperative Multi-agent Reinforcement Learning in Open Environments}

The previous content mainly discussed in the context of classical closed environments, where the factors in the environment remain constant. However, research on machine learning algorithms in real environments often needs to address situations where certain factors may change. This characteristic has given rise to a new research area—open-environment machine learning, including Open-world Learning~\cite{conole2012designing}, Open-environment Learning~\cite{zhou2022open}, Open-ended Learning~\cite{song2022little}, Open-set Learning~\cite{geng2020recent}, etc.

\subsection{Machine Learning in Open Environments}
Traditional machine learning is generally discussed in classical closed environments, where important factors in the environment do not change. These factors can have different definitions and scopes in different research fields, such as new categories in supervised learning, new tasks in continual learning, changes in features in neural network inputs, shifts in the distribution of training (testing) data, and changes in learning objectives across tasks. Open-environment machine learning~\cite{zhou2022open}, to some extent, is related to open-ended learning~\cite{song2022little}. In contrast to machine learning in closed environments, open-environment machine learning primarily considers the possibility of changes in important factors in the machine learning environment~\cite{zhou2022open}. Previous research in this area has explored category changes in supervised learning~\cite{parmar2023open}, feature changes~\cite{hou2019learning}, task changes in continual learning~\cite{kim2023open}, setting of open environments for testing~\cite{grbic2021evocraft}, and open research in game problems~\cite{balduzzi2019open}, among others.

On the other hand, open-environment RL has gained attention and made progress in various aspects in recent years. Unlike supervised learning, where model is learned from labeled training data, or unsupervised learning, which analyzes inherent information in given data, RL places agents in unfamiliar environments. Agents must interact autonomously with the environment and learn from the results of these interactions. Different interaction methods generate different data, posing a key challenge in RL—learning from dynamic data distributions that, in turn, further alter the data distribution~\cite{sutton2018reinforcement}. In addition, agents need to consider changes in the Markov Decision Process (MDP) of the environment. Some methods  have attempted to learn a trustworthy RL policy~\cite{Xu2022TrustworthyRL}, learn a highly generalizable policy through methods like evolutionary learning~\cite{wang2020enhanced}, learn skills for open environments~\cite{yuan2023plan4mc}, enhance generalization ability by changing the reward function~\cite{meier2022open}, design testing environments for open RL~\cite{matthews2022skillhack}, design general open intelligent agents based on RL~\cite{team2021open,team2023human}, study policy generalization in RL~\cite{kirk2023survey}, etc.

\subsection{Cooperative Multi-Agent Reinforcement Learning in Open Environments}

\begin{table*}
\renewcommand{\arraystretch}{1}
    \centering
    \caption{Research Directions in Multi-Agent Systems in Open Environments.}
    \resizebox{0.94\textwidth}{!}{
    \begin{tabularx}{\textwidth}{ZZYZ}
        \toprule
        Research Direction & Core Content & Representative Algorithms & Applications and Achievements\\
        \midrule
        \vtop{\hbox{\strut Offline MARL}} & Extending successful offline learning techniques from single-agent RL to multi-agent scenarios or designing multi-agent offline methods specifically & ICQ~\cite{icq}, MABCQ~\cite{mabcq}, SIT~\cite{tian2022learning}, ODIS~\cite{zhang2023discovering}, MADT~\cite{meng2021offline}, OMAC~\cite{wang2023offline}, CFCQL~\cite{shao2023counterfactual} & Learning policies from collected static offline data, avoiding issues arising from interaction with the environment, achieving learning objectives from large-scale and diverse data\\ 
        \midrule
        \vtop{\hbox{\strut Policy Transfer}\hbox{\strut and Generalization}} & Transferring and directly generalizing multi-agent policies across tasks to enable knowledge reuse & LeCTR~\cite{DBLP:conf/aaai/OmidshafieiKLTR19}, MAPTF~\cite{yang2021efficient},
        EPC~\cite{long2019evolutionary}, Literature~\cite{mahajan2022generalization}, MATTAR~\cite{qin2022multi}  & Facilitating knowledge reuse across tasks, speeding up learning on new tasks\\
        \midrule
        
        \vtop{\hbox{\strut Continual}\hbox{\strut Cooperation}} & Cooperative task learning when facing tasks or samples presented sequentially & Literature
        ~\cite{nekoei2021continuous}, MACPro~\cite{yuan2023multimacpro}, Macop~\cite{yuan2023learningmacop} & Extending existing techniques from single-agent scenarios to handle cooperative task emergence in multi-agent settings\\
        \midrule
        \vtop{\hbox{\strut Evolutionary}\hbox{\strut MARL}} & Simulating heuristic stochastic optimization algorithms inspired by the natural evolution process, including genetic algorithms, evolutionary policies, particle swarm algorithms, etc., to empower multi-agent coordination & MERL~\cite{majumdar2020evolutionary}, BEHT~\cite{dixit2022balancing}, MCAA~\cite{dixit2022diversifying}, EPC~\cite{long2019evolutionary}, ROMANCE~\cite{yuan2023robustromance}, MA3C~\cite{yuan2023communication} & Simulating multi-agent policies through evolutionary algorithms or generating training partners or opponents to assist multi-agent policy training, widely applied in various project scenarios\\
        \midrule    
        \vtop{\hbox{\strut Robustness}\hbox{\strut in MARL}} & Considering policy learning and execution when the system environment undergoes changes, learning robust policies to cope with environmental noise, teammate changes, etc. & R-MADDPG~\cite{zhang2020robustmarl}, Literature~\cite{van2020robust}, RAMAO~\cite{wang2022robust}, ROMANCE~\cite{yuan2023robustromance}, MA3C~\cite{yuan2023comrobust}, CroMAC~\cite{yuan2023communication} & Maintaining robust cooperative capabilities in conditions where states, observations, actions, and communication channels in the environment are subjected to noise or even malicious attacks\\ 
        \midrule
        \vtop{\hbox{\strut Multi-Objective}\hbox{\strut (Constrained)}\hbox{\strut Cooperation}} & Optimizing problems with multiple objectives, simultaneously considering the optimal solutions of different objective functions & MACPO(MAPPO-Lagrangian)~\cite{gu2023safe}, CAMA~\cite{wang2022cama}, MDBC~\cite{qin2020learning}, Literature~\cite{elsayed2021safe,melcer2022shield,xiao2023model} & Addressing multiple constraint objectives in the environment, making progress in constrained or safety-related domains, laying the foundation for practical applications of multi-agent coordination\\
        \midrule
        \vtop{\hbox{\strut Risk-Sensitive}\hbox{\strut MARL}} & Modeling the numerical values (rewards) of variables in the environment as distributions using value distributions, using risk functions to assess system risks, etc. & DFAC~\cite{sun2021dfac}, RMIX~\cite{qiu2021rmix}, ROE~\cite{oh2023toward}, DRE-MARL~\cite{hu2022distributional}, DRIMA~\cite{son2022disentangling}, Literature~\cite{liu2023learning,slumbers2023game} & Enhancing coordination performance in complex scenarios, effectively perceiving risks, and assessing performance in risk-sensitive scenarios\\
        \bottomrule
    \end{tabularx}}
   \label{opendmarl}
\end{table*}

\begin{table*}[h]
\renewcommand{\arraystretch}{1}
    \centering
    \caption{Research Directions in Multi-Agent Systems in Open Environments (Continued).}
    \resizebox{0.94\textwidth}{!}{
    \begin{tabularx}{\textwidth}{ZZYZ}
        \toprule
        Research Direction & Core Content & Representative Algorithms & Applications and Achievements\\
        \midrule
        \vtop{\hbox{\strut Ad-hoc}\hbox{\strut Teamwork}} & Creating a single autonomous agent capable of efficiently and robustly collaborating with unknown teammates in specific tasks or situations & Literature~\cite{DBLP:conf/aaai/StoneKKR10,Chandrasekaran2016IndividualPI}, ODITS~\cite{DBLP:conf/iclr/GuZH022}, OSBG~\cite{rahman2021towards}, BRDiv~\cite{rahman2023generating}, L-BRDiv~\cite{rahman2023minimum}, TEAMSTER~\cite{ribeiro2023teamster} & Empowering a single autonomous agent with the ability to collaborate rapidly with unknown teammates, achieving quick coordination in temporary teams across various task scenarios\\
        \midrule
        \vtop{\hbox{\strut Zero (Few)-shot}\hbox{\strut Coordination}} & Designing training paradigms to enable MASs to collaborate with unseen teammates using few or zero interaction samples & FCP~\cite{strouse2021collaborating}, TrajeDi~\cite{lupu2021trajectory}, MAZE~\cite{xue2022heterogeneous}, CSP~\cite{ding2023coordination}, LIPO~\cite{lipo}, HSP ~\cite{yu2023learning}, Macop~\cite{yuan2023multimacpro}, Literature~\cite{nekoei2023towards,fosong2022few} & Current algorithms have shown effective zero-shot or low-shot coordination with diverse unseen teammates in benchmark environments like Overcooked \\
        \midrule
        \vtop{\hbox{\strut Human-AI}\hbox{\strut Coordination}} & Providing support for human-intelligence (huma-machine) interaction, enabling better coordination between human participants and agents to accomplish specific tasks & FCP~\cite{strouse2021collaborating}, Literature~\cite{hu2022human}, HSP~\cite{yu2023learning}, Latent Offline RL~\cite{hong2023learning}, RILI~\cite{parekh2023learning}, PECAN~\cite{lou2023pecan} & Achieving certain levels of coordination between human and machine in given simulation environments or real-world robot scenarios \\
        \midrule
        \vtop{\hbox{\strut Cooperative}\hbox{\strut Large Language Models}} & Developing large cooperative decision-making models using the concept of universal large language models, or leveraging current large language model technologies to enhance multi-agent coordination & MADT~\cite{meng2021offline}, MAT~\cite{wen2022multiagent}, MAGENTA~\cite{wang2023multiagent}, MADiff~\cite{zhu2023madiff}, ProAgent~\cite{zhang2013coordinating}, SAMA~\cite{li2023semantically} & For specific task scenarios, using large language model to impart policies with a degree of generality; additionally, some works leverage large language models to promote system coordination capabilities\\  
        \bottomrule
    \end{tabularx}}
   \label{opendmarlcon}
\end{table*}

The aforementioned content introduced some related works on SARL in open environments. Some studies have provided descriptions of MARL in open environments from specific perspectives. In open MASs, the composition and scale of the system may change over time due to agents joining or leaving during the coordination process~\cite{hendrickx2017open}.

Classical MARL algorithms primarily address issues such as non-stationarity caused by teammate policy optimization during training and the exploration and discovery of effective coordination patterns. While these methods effectively improve sample efficiency and cooperative ability, they do not consider the problem of changes in real-world MASs and environmental factors. Taking the example of open MASs mentioned earlier, due to changes in teammate behavior styles, the intelligent agents generated by classical MARL algorithms make decisions based on historical information. They cannot promptly perceive changes in teammate behavior styles, leading to a certain lag in their adaptive capability, greatly impacting cooperative performance.

In previous works, research focused mainly on multi-agent planning in open situations, giving rise to many related problem settings, such as Open Decentralized Partially Observable Markov Decision Processes (Open Dec-POMDP)~\cite{Cohen2017OpenDP}, Team-POMDP~\cite{Cohen2018MonteCarloPF,Cohen2019PowerIF}, I-POMDP-Lite~\cite{Chandrasekaran2016IndividualPI,Eck2019ScalableDP}, and CI-POMDP~\cite{Kakarlapudi2022DecisiontheoreticPW}. Recently, some work has begun to consider the issue of open MARL. GPL~\cite{rahman2021towards} formalizes the problem of Open Ad-hoc Teamwork as Open Stochastic Bayesian Games (OSBG), assuming global observability to improve efficiency. However, it is challenging to implement in the real world. Additionally, it uses a graph neural network-based method applicable only to single-controllable intelligent agent settings, making it challenging to extend to multiple controllable intelligent agents. Recent work~\cite{eck2023decision} proposes OASYS, an open MAS, to describe open MASs.

While the aforementioned works have delved into research on MARL in open environments from specific perspectives, their focus is relatively narrow, and there are certain limitations. There lacks a comprehensive overview of the entire research field. We believe that for a cooperative MAS to be applied to complex open real-world scenarios, it should possess the ability to cope with changes in environmental factors (states, actions, reward functions, etc.), changes in coordination modes (teammates, opponents), and the emergence of tasks in the form of data streams. This capability should mainly include the following aspects:

\begin{itemize}
    \item During policy training and evolution, the system should have the ability for offline policy learning, policies with transfer and generalization capabilities, policies that support continual learning, and the system should possess evolution and adaptation capabilities.
    \item In the deployment process, policies should be able to handle changes in environmental factors, specifically demonstrating robust cooperative capabilities when states, observations, actions, environmental dynamic, and communication change in environments.
    \item In real-world deployment, considerations should include multi-objective (constraint) policy optimization, risk perception and assessment capabilities in the face of real, high-dynamic task scenarios.
    \item Trained policies, when deployed, should have self-organizing cooperative capabilities and should have zero (or few) shot adaptation capabilities. Furthermore, they should support human-intelligence coordination, endowing the MAS with the ability to serve humans.
    \item Finally, considering the differences and similarities in various multi-agent cooperative tasks, learning policy models for each type of task often incurs high costs and resource waste. Policies should have the capability o cover various multi-agent cooperative tasks, similar to ChatGPT.
\end{itemize}

Based on this, this section reviews and compares relevant works in these eleven aspects, presenting the main content and existing issues in current research, as well as future directions worthy of exploration.

\subsubsection{Offline Cooperative Multi-Agent Reinforcement Learning}
Offline reinforcement learning~\cite{levine2020offline,prudencio2023survey} has recently attracted considerable research attention, focusing on a data-driven training paradigm that does not require interaction with the environment~\cite{prudencio2023survey}. Previous work~\cite{bcq} primarily addressed distributional shift issues in offline learning, considering learning behavior-constrained policies to alleviate extrapolation errors in estimating unseen data~\cite{brac, bear, cql}. Offline MARL is a relatively new and promising research direction~\cite{DBLP:journals/tac/ZhangYLZB21}, training cooperative policie's from static datasets. One class of offline MARL methods attempts to learn policies from offline data with policy constraints. ICQ~\cite{icq} effectively mitigated extrapolation errors in MARL by trusting only offline data. MABCQ~\cite{mabcq} introduced a fully distributed setting for offline MARL, utilizing techniques such as value bias and transfer normalization for efficient learning. OMAR~\cite{omar} combined first-order policy gradients and zero-order optimization methods to avoid discordant local optima. MADT~\cite{meng2021offline} harnessed the powerful sequential modeling capability of Transformers, seamlessly integrating it with offline and online MARL tasks. Literature~\cite{tian2022learning} investigated offline MARL, explicitly considering the diversity of agent trajectories and proposing a new framework called Shared Individual Trajectories (SIT). Literature~\cite{tseng2022offline} proposes training a teacher policy with access to observations, actions, and rewards for each agent first. After determining and gathering "good" behaviors in the dataset, individual student policies are created and endowed with the structural relationships between the teacher policy's features and the agents through knowledge distillation. ODIS~\cite{zhang2023discovering} introduced a novel offline MARL algorithm for discovering cooperative skills from multi-task data. Literature~\cite{formanek2023off} recently released the Off-the-Grid MARL (OG-MARL) framework for generating offline MARL datasets and algorithm evaluation. M3~\cite{meng2023m3} innovatively introduced the idea of multi-task and multi-agent offline pre-training modules to learn higher-level transferable policy representations. OMAC~\cite{wang2023offline} proposed an offline MARL algorithm based on coupled value decomposition, decomposing the global value function into local and shared components while maintaining credit assignment consistency between global state values and Q-value functions.

\subsubsection{Cooperative Policy Transfer and Generalization}
Transfer learning is considered a crucial method to improve the sample efficiency of RL algorithms~\cite{zhu2023transfer}, aiming to reuse knowledge across different tasks, accelerating the policy learning of agents on new tasks. Transfer learning in multi-agent scenarios~\cite{da2019survey} has also garnered extensive attention. In addition to considering knowledge reuse between tasks, some researchers focus on knowledge reuse among agents. The basic idea is to enable some agents to selectively reuse knowledge from other agents, thereby assisting the overall MAS in achieving better cooperation. DVM~\cite{DBLP:conf/iros/WadhwaniaKOH19} modeled the multi-agent problem as a multi-task learning problem, combining knowledge between different tasks and distilling this knowledge through a value-matching mechanism. LeCTR~\cite{DBLP:conf/aaai/OmidshafieiKLTR19} conducted policy teaching in multi-agent scenarios, guiding some agents to guide others, thus facilitating better policy cooperation overall. MAPTF~\cite{yang2021efficient} proposed an option-based policy transfer method to assist multi-agent cooperation.

On the other hand, methods for policy reuse among multi-agent tasks emphasize the reuse of knowledge and experience from old tasks to assist in the policy learning of new tasks, focusing on knowledge transfer between different tasks. Compared to single-agent problems, the dimensions of environmental state (observations) may vary in tasks with different scales of multi-agent tasks, posing challenges to policy transfer among tasks. Specifically, DyMA-CL~\cite{wang2020few} designed a network structure independent of the number of agents and introduces a series of transfer mechanisms based on curriculum learning to accelerate the learning process of cooperative policies in multi-agent scenarios. EPC~\cite{long2019evolutionary} proposed an evolutionary algorithm-based multi-agent curriculum learning method to help groups achieve cooperative policy learning in complex scenarios. UPDeT~\cite{hu2021updet} and PIT~\cite{zhou2021cooperative} leveraged the generalization of transformer networks to address the problem of changing environmental input dimensions, aiding efficient cooperation and knowledge transfer among agent groups. These related works on multi-agent transfer learning provide inspiration for knowledge transfer between tasks, but they do not explicitly consider the correlation between tasks and how to
utilize task correlation for more efficient knowledge transfer remains an open research topic. MATTAR~\cite{qin2022multi} addressed the challenge of adapting cooperative policy models to new tasks and proposes a policy transfer method based on task relations. Literature~\cite{shi2023lateral} considers using lateral transfer learning to promote MARL. Literature~\cite{mahajan2022generalization} further focused on designing algorithms to enhance the generalization ability in MARL.

\begin{figure*}
\centering
\includegraphics[width=0.7\linewidth]{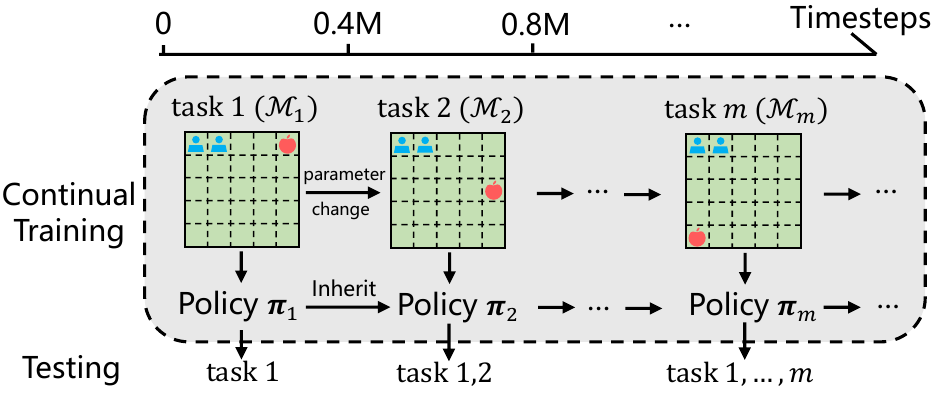}
 \caption{Illustration of continual coordination.}
  \label{contnualtoy}
\end{figure*}

\subsubsection{Multi-Agent Reinforcement Learning and Continual coordination}
Continual learning, incremental learning and lifelong learning are related, assuming tasks or samples appear sequentially~\cite{kudithipudi2022biological}. In recent years, continual RL~\cite{khetarpal2022towards,abel2023definition} has received some attention. In this setting, the challenge for agents is to avoid catastrophic forgetting while transferring knowledge from old tasks to new ones (also known as the stability-plasticity dilemma~\cite{parisi2019continual}), while maintaining scalability to a large number of tasks. Researchers have proposed various methods to address these challenges. EWC~\cite{kirkpatrick2017overcoming} used weight regularization based on $l_2$ distance to constrain the gap between current model parameters and those of previously learned models, requiring additional supervision to select specific Q-function heads and set exploration policies for different task scenarios. CLEAR~\cite{Rolnick2018ExperienceRF} is a task-agnostic continual learning method that does not require task information during the continual learning process. It maintains a large experience replay buffer and addresses forgetting by sampling data from past tasks. Other methods such as HyperCRL~\cite{huang2021continual} and~\cite{kessler2022surprising} leveraged learned world models to enhance learning efficiency. To address scalability issues in scenarios with numerous tasks, LLIRL~\cite{DBLP:journals/tnn/WangCD22} decomposed the task space into subsets and uses a Chinese restaurant process to expand neural networks, making continual RL more efficient. OWL~\cite{DBLP:conf/aaai/KesslerPBZR22} is a recent efficient method based on a multi-head architecture. CSP~\cite{gaya2023building} gradually constructed policy subspaces, training RL agents on a series of tasks. Another class of means to address scalability issues is based on the idea of Packnet~\cite{mallya2018packnet}, sequentially encoding task information into neural networks and pruning network nodes for relevant tasks. Regarding the problem of continual learning in multi-agent settings (see Figure~\ref{contnualtoy}), literature~\cite{nekoei2021continuous} investigated whether agents can cooperate with unknown agents by introducing a multi-agent learning testbed based on Hanabi. However, it only considers single-modal task scenarios. MACPro~\cite{yuan2023multimacpro} proposed a method for achieving continual coordination among multiple agents through progressive task contextualization. It uses a shared feature extraction layer to obtain task features but employs independent policy output heads, each making decisions for tasks of specific categories. Macop~\cite{yuan2023learningmacop} endowed a MAS with continual coordination capabilities, developing an algorithm based on incompatible teammate evolution and high-compatibility multi-agent cooperative training paradigms. In various test environments where teammates switch within and between rounds, the proposed method can quickly capture teammate identities, demonstrating stronger adaptability and generalization capabilities compared to various comparative methods, even in unknown scenarios.

\subsubsection{Evolutionary Multi-Agent Reinforcement Learning}
\begin{figure*} 
\centering
\includegraphics[width=0.8\linewidth]{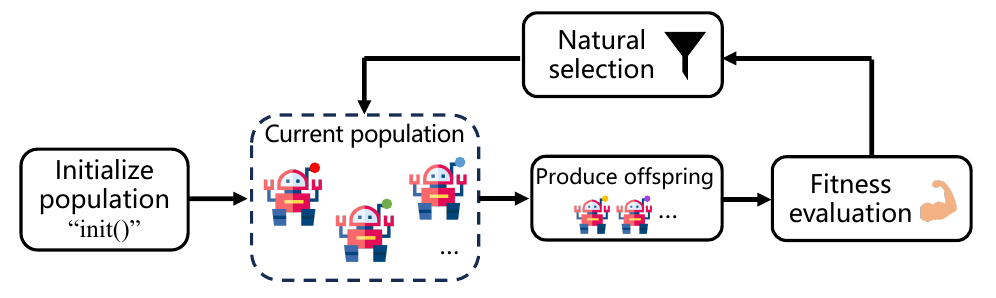}
\caption{Illustration of Evolutionary Algorithms.}\label{evfig}
\end{figure*}
Evolutionary algorithms~\cite{zhou2019evolutionary} constitute a class of heuristic stochastic optimization algorithms simulating the natural evolution process, including genetic algorithms, evolutionary policies, particle swarm algorithms, etc. (Figure~\ref{evfig}). Despite numerous variants, the main idea remains consistent in solving problems. First, a population is initialized by generating several individuals as the initial population through random sampling. The remaining process can be abstracted into a loop of three main steps. Based on the current population, offspring individuals are generated using operators such as crossover and mutation; the fitness of offspring individuals is evaluated; according to the survival of the fittest principle, some individuals are eliminated, and the remaining individuals constitute the new generation. Previous research~\cite{DBLP:conf/nips/QianYZ15} revealed the rich potential of evolutionary algorithms in solving subset selection problems. Evolutionary algorithms have also found widespread applications in the multi-agent domain~\cite{bloembergen2015evolutionary}, such as $\alpha$-Rank~\cite{omidshafiei2019alpha} and many subsequent works that used evolutionary algorithms to optimize and evaluate MASs.

In the context of cooperative tasks, evolutionary algorithms play a significant role. Literature~\cite{shibata1994coordination} considered the distributed configuration problem in multi-agent robot systems, optimizing the system through a fuzzy system and evolutionary algorithm to enhance cooperative performance. MERL~\cite{majumdar2020evolutionary} designed a layered training platform, addressing two objectives through two optimization processes. An evolutionary algorithm maximizes sparse team-based objectives by evolving the team population. Simultaneously, a gradient-based optimizer trains policies, maximizing rewards for specific individual agents. BEHT~\cite{dixit2022balancing} introduced high-quality diverse goals into MASs to solve heterogeneous problems, effectively enhancing system generalization. MCAA~\cite{dixit2022diversifying} and some subsequent works~\cite{li2023race,dixit2023learning} considered improving asymmetric MASs through evolutionary learning. EPC~\cite{long2019evolutionary} enhanced the generalization and transfer capabilities of MASs through population evolution. ROMANCE~\cite{yuan2023robustromance} and MA3C~\cite{yuan2023communication} used population evolution to generate adversarial attackers to assist in the training of cooperative MASs, obtaining robust policies.

\subsubsection{Robust Cooperation in Multi-Agent Reinforcement Learning}
The study of robustness in RL has garnered widespread attention in recent years and has made significant progress~\cite{moos2022robust}. The focus of robustness research includes perturbations to various aspects of agents in RL, such as states, rewards, and actions. One category of methods introduces auxiliary adversarial attackers, achieving robustness through adversarial training alternating between the policy in use and the opponent system~\cite{pinto2017robust,vinitsky2020robust,zhang2020robust,song2022robust}. Other methods enhance robustness by designing appropriate regularization terms in the loss function~\cite{oikarinen2021robust,sun2021strongest,liangefficient}. In comparison to adversarial training, these methods effectively improve sample efficiency. However, these approaches lack certifiable robustness guarantees regarding the noise level and the robustness between policy execution. In response to this issue, several verifiable robustness methods have been derived~\cite{qin2020learning,everett2021certifiable, wu2021crop,wu2021copa}.

Currently, research on robustness in MARL has started to attract attention~\cite{guo2022towards}. The main challenges lie in the additional considerations required for MASs compared to single-agent systems, including non-stationarity arising from complex interactions among agents~\cite{papoudakis2019dealing}, trust allocation~\cite{wang2021towards}, scalability~\cite{christianos2021scaling}, etc. Early work aimed to investigate whether cooperative policies exhibit robustness. For instance, targeting a cooperation policy trained using the QMIX algorithm, literature~\cite{lin2020robustness} trained an attacker against observations using RL. Subsequently, literature~\cite{guo2022towards} conducted a comprehensive robustness test on typical MARL algorithms such as QMIX and MAPPO in terms of rewards, states, and actions. The results further confirmed the vulnerability of MASs to attacks, emphasizing the necessity and urgency of research on robust MARL. Recent progress in enhancing robustness in MARL includes work focusing on learning robust cooperative policies to avoid overfitting to specific teammates~\cite{van2020robust} and opponents~\cite{li2019robust}. Similar to robust SARL, R-MADDPG~\cite{zhang2020robustmarl} addressed the uncertainty of the MAS's model, establishing the concept of robust Nash equilibrium under model uncertainty and achieving optimal robustness in multiple environments. Addressing the issue of perturbed actions in some agents within MASs, literature~\cite{hu2021robust} introduces a heuristic rule and relevant equilibrium theory to learn a robust cooperative policy for MASs. The robustness of multi-agent communication has also received attention in recent years. Literature~\cite{mitchell2020gaussian} designed a filter based on Gaussian processes to extract valuable information from noisy communication. Literature~\cite{tu2021adversarial} studied the robustness of multi-agent communication systems at the neural network level. Literature~\cite{xue2022mis} modeled multi-agent communication as a two-player zero-sum game and applies PSRO technology to learn a robust communication policy. Works such as ARTS~\cite{phan2020learning} and RADAR~\cite{phan2021resilient} considered the resilience of MASs and study the recovery capabilities of cooperative MARL tasks in the face of environmental changes.

Recently, addressing the challenge of cooperation robustness in dynamically changing environments, literature~\cite{yuan2023robustromance} proposes the ROMANCE algorithm, an  robust cooperation algorithm against evolutionary auxiliary adversarial attackers. Literature~\cite{yuan2023comrobust} introduces the MA3C framework for adversarial robust multi-agent communication training through population adversarial training. Literature~\cite{yuan2023communication} proposes the CroMAC method for verifiable robust communication with a multi-view information perspective. Research on perturbed multi-agent observations is discussed in literature~\cite{wang2022robust}, and literature~\cite{he2023robust} considers learning a robust policy against state attacks in MASs.

\subsubsection{Multi-Objective (Constrained) Cooperative Multi-Agent Reinforcement Learning}
Multi-objective optimization~\cite{deb2016multi} refers to the existence of multiple objective functions in an optimization problem, requiring simultaneous consideration of the optimal solutions for each objective. In multi-objective optimization problems, conflicts may arise between different objective functions, meaning that improving one objective function may lead to the deterioration of another. Therefore, a balance needs to be struck between different objective functions to find a compromise point, also known as the Pareto optimal solution set.

In RL, multi-objective optimization is frequently employed. For example, in multi-objective RL, agents need to learn Pareto optimal policies across multiple objectives~\cite{roijers2013survey,liu2014multiobjective,DBLP:conf/nips/YangSN19,hayes2022practical}. Similarly, in MARL, some works introduce multi-objective learning problems, generally modeled as Multiobjective MASs (MOMAS)~\cite{ruadulescu2020multi}, where reward functions for different objectives may conflict. Literature~\cite{DBLP:conf/ijcai/DurugkarLS20} considered the relationship between individual preferences and shared goals in MASs, modeling it as a multi-objective problem. The results indicate that a mixed processing approach achieves better performance than solely considering a single objective. Literature~\cite{ropke2023reinforcement} exploreed how communication and commitment help multi-agents learn appropriate policies in challenging environments. Literature~\cite{DBLP:journals/nca/RadulescuVZMRN22} addressed the multi-objective opponent modeling problem in general game problems, accelerating policy learning through multi-objectives.

On another note, recent works focus on single-agent safe RL~\cite{garcia2015comprehensive,gu2022review} and multi-agent safe RL~\cite{gu2023safe}. These constrained problems in RL settings can be modeled as Constrained Markov Decision Processes (CMDP). In MARL, literature~\cite{gu2023safe} proposed testing environments, Safe MAMuJoCo, Safe MARobosuite, and Safe MAIG, for multi-agent tasks. They further propose safe MARL algorithms MACPO and MAPPO-Lagrangian. Literature~\cite{ding2023provably} studied online secure MARL in constrained Markov games, where agents compete by maximizing their expected total utility while constraining the expected total reward. Literature~\cite{ying2023scalable} investigated safe MARL, where agents attempt to jointly maximize the sum of local objectives while satisfying their individual safety constraints. CAMA~\cite{wang2022cama} explored safety issues in multi-agent coordination. Literature~\cite{zhang2023safe} considered robustness and safety issues in MASs when states are perturbed. Additionally, some works consider safety in
MARL based on barrier protection~\cite{elsayed2021safe,melcer2022shield,xiao2023model}, or combine safety in MASs with control techniques~\cite{qin2020learning,dawson2023safe}.

\subsubsection{Risk-Aware Multi-Agent Reinforcement Learning}
Distributional RL has made significant progress in various domains in recent years~\cite{bellemare2023distributional}. Classical value-based RL methods attempt to model the cumulative return using expected values, represented as value functions $V(s)$ or action-value functions $Q(s,a)$. However, complete distribution information is largely lost in this modeling process. Distributional RL aims to address this issue by modeling the distribution $Z(s,a)$ of the random variable representing the cumulative return. Such methods are also applied to multi-agent cooperation tasks. To alleviate the randomness in the environment caused by local observability, DFAC~\cite{sun2021dfac} extended the reward function of a single agent from a deterministic variable to a random variable. It models the mixing function of QMIX-type algorithms as a distributional mixing function, achieving excellent cooperation results in various challenging tasks. Furthermore, to mitigate the uncertainty introduced by the stochasticity of reward functions in multi-agent cooperation tasks, RMIX~\cite{qiu2021rmix} utilized risk-based distributional techniques, such as Conditional Value at Risk (CVaR), to enhance the algorithm's cooperation ability. The algorithm innovatively introduces risk assessment based on the similarity of agent trajectories, with theoretical justification and experimental results verifying its effectiveness. ROE~\cite{oh2023toward} proposed a risk-based optimistic exploration method from another perspective. This method selectively samples distributions to effectively improve the exploration efficiency of MASs.

In addition to distribution-based multi-agent cooperation algorithms mentioned above, other works explore various aspects, such as reward evaluation based on value distribution~\cite{hu2022distributional}, efficient and adaptive multi-agent policy learning in general game problems~\cite{liu2023learning}, risk decoupling learning in the multi-agent learning process~\cite{son2022disentangling}, and risk management based on game theory~\cite{slumbers2023game}. Although these works have achieved certain results in various environments, considering the unknown risks in real environments, exploring how to deploy multi-agent policies in real environments, automatically identifying environmental risks, and adjusting cooperative policies accordingly is one of the future research directions.

\subsubsection{Ad-hoc Teamwork}
Ad-hoc teamwork (AHT)~\cite{mirsky2022survey} aims to endow intelligent agents with the ability to efficiently collaborate with untrained agents, creating autonomous agents capable of effective and robust coordination with previously unknown teammates~\cite{DBLP:conf/aaai/StoneKKR10}. Early work assumed knowledge of the cooperative behavior of teammates for learning agents~\cite{DBLP:conf/aaai/StoneKKR10,DBLP:conf/aamas/AgmonS12}. Subsequent research gradually relaxed this assumption, allowing autonomous learning agents to be unaware of teammates' behaviors during interaction. Some approaches design algorithms to predict corresponding teammates' policies by observing their behaviors, promoting coordination in AHT~\cite{DBLP:conf/atal/AlbrechtS17,barrett2017making,DBLP:conf/ijcai/RavulaAS19,macke2021expected}. Other works attempted to enhance cooperation among teammates in AHT through effective communication methods\cite{DBLP:conf/atal/BarrettAHKS14}. While these approaches improve coordination performance to some extent, they assume that cooperating teammates are in a closed environment, maintaining a constant number and type of teammates in a single trajectory. Open AHT has been proposed and studied to address this limitation~\cite{Chandrasekaran2016IndividualPI}, where GPL~\cite{rahman2021towards} tackles changes in teammate types and quantities at different time points through a graph neural network.

Early AHT work generally considered agents in globally observable environments. Recent efforts extend this setting to scenarios with locally observable scenes. ODITS~\cite{DBLP:conf/iclr/GuZH022} evaluated other teammates' behaviors using mutual information-optimized regularization, enabling trained autonomous agents to infer teammate behavior from local observations. In contrast to previous studies, a method is proposed to address open-ended temporary teamwork in partially observable scenarios~\cite{rahman2022general}. TEAMSTER ~\cite{ribeiro2023teamster} introduced a method to decouple world model learning from teammate behavior model learning. Additionally, some work explores various aspects, including AHT problems with attackers~\cite{fujimoto2022ad}, few-shot interaction coordination~\cite{fosong2022few}, and teammate generation coverage in AHT~\cite{rahman2023generating,rahman2023minimum} .

\subsubsection{Zero (Few)-Shot Coordination}
Zero-shot coordination (ZSC) is a concept proposed in recent years for cooperative multi-agent tasks, aiming to train agents to cooperate with unseen teammates~\cite{treutlein2021new}. Self-play~\cite{tesauro1994td, silver2018general} is one means of effectively improving coordination, where agents continuously enhance their coordination abilities through self-coordination. However, agents generated in this way may lack the ability to collaborate with unseen teammates. Literature ~\cite{hu2020other} further refined the problem, involving introducing sequence-independent training methods to alleviate suboptimality. To address potential overfitting to specific teammate behavior styles resulting from training with a single teammate, other methods, such as Fictitious Co-Play (FCP)~\cite{heinrich2015fictitious, strouse2021collaborating} and co-evolution with teammate populations~\cite{xue2022heterogeneous}, have achieved success. Some works leverage few-shot techniques to cope with multimodal scenarios and have shown effectiveness~\cite{ding2023coordination, yuan2023multimacpro}. Recent research~\cite{wang2023quantifying} evaluates and quantifies the capacity of various ZSC algorithms based on the action preferences of collaborating teammates.

Beyond these efforts, ZSC research includes diversity metrics~\cite{lupu2021trajectory,zhao2021maximum}, design of training paradigms~\cite{hu2020other,strouse2021collaborating,xue2022heterogeneous}, equivariant network design~\cite{muglich2022equivariant}, coordination enhancement based on policy similarity evaluation~\cite{yu2023improving}, general scenarios of ZSC problems~\cite{li2023cooperative}, ZSC improvement based on ensemble techniques~\cite{li2023cooperative}, studies on human value preferences~\cite{yu2023learning}, diverse teammate generation~\cite{lipo}, and policy coevolution in heterogeneous environments~\cite{xue2022heterogeneous}. Additionally, few-shot adaptation has found widespread use in single-agent meta-RL~\cite{kumar2020one,osa2022discovering,gaya2022learning}, and Few-shot Teamwork (FST)~\cite{fosong2022few} explores generating agents that can adapt and collaborate in unknown but relevant tasks. CSP~\cite{ding2023coordination} considered a multi-modal cooperation paradigm and develops a few-shot collaboration paradigm that decouples collaboration and exploration strategies, it collects a small number of samples during policy execution to find the optimal policy head. The study in~\cite{nekoei2023towards} found that current performance-competitive ZSC algorithms require a considerable number of samples to adapt to new teammates when facing different learning methods. Accordingly, they propose a few-shot collaborative method and validate the algorithm's effectiveness on Hanabi. Macop~\cite{yuan2023learningmacop} considered the adaptability of strategies under changes in cooperative objects between rounds and proposes a highly compatible collaboration algorithm for arbitrary teammates, significantly improving the collaboration algorithm's generalization ability, showing astonishing coordination effectiveness.

\subsubsection{Human-AI Coordination}
The capability to enable efficient collaboration between agents (robots) and humans has been a longstanding goal in artificial intelligence~\cite{ajoudani2018progress, vicentini2021collaborative}. The concept of human-AI coordination~\cite{DBLP:conf/nips/CarrollSHGSAD19} has a relationship with Human-AI Interaction (HAI)~\cite{van2021human} or Human-Robot Interaction (HRI)~\cite{onnasch2021taxonomy}. The purpose of human-AI coordination is to enhance collaboration between human participants and agents to accomplish specific tasks. Benefiting the strong problem solving ability, cooperative MARL can be employed to improve human-AI coordination capabilities for different human participants.

In contrast to the previously mentioned ZSC issue, human-AI coordination considers human participants as cooperative entities. Although research has shown that, in some environments, it might be possible for intelligent agents to collaborate with real humans without training on human data~\cite{strouse2021collaborating}, in scenarios where subtle features of human behavior critically impact the task, generating effective collaborative strategies is impossible without human data. For the intelligent agents to be trained, one approach is to directly encode teammates with human behavioral styles through prior bias~\cite{hu2020other, wang2020too, puig2020watch}. Another approach involves training the agents to varying degrees using data collected from interactions with real humans. Some methods combine hand-coding human behaviors based on prior bias with optimizing agents using data from human interactions~\cite{lerer2019learning, tucker2020adversarially, shih2020critical}.

However, these methods make strong assumptions about the patterns of human behavior during testing, which are often unrealistic. In response to this issue, a series of methods have emerged to learn models of human behavior and compute the best responses to them, facilitating human-AI coordination~\cite{hu2022human}. In the realm of human-aided collaboration, some methods are exploring alternative perspectives, such as studying task scenarios preferred by humans~\cite{yu2023learning}, promoting human-machine coordination through offline data~\cite{hong2023learning}, developing technology for human-AI mutual cooperation~\cite{parekh2023learning}, exploring leadership and following technologies in human-machine coordination~\cite{li2021influencing}, zero-shot human-AI coordination~\cite{li2023tackling}, Bayesian optimization-based human-AI coordination~\cite{av2022human}, and setting up human-machine collaboration environments~\cite{thumm2023humanrobot}. Although these works have made some progress in human-aided collaboration, several challenges persist in this direction. For instance, there is a lack of convenient and effective testing environments; most work is primarily conducted on Overcooked~\cite{DBLP:conf/nips/CarrollSHGSAD19}, which has a limited number of agents and overly simple scenarios. Additionally, some studies mainly validate in third-party non-open-source environments like custom-made robotic arms, posing challenges in developing versatile testing environments suitable for various task requirements and human participants, as well as designing more efficient algorithms. On the other hand, approaches such as human value alignment~\cite{hussonnois2023controlled, yuan2022situ}, and human in-the-loop training~\cite{guo2022interactive} could be potential solutions to address these issues in the future.

\subsubsection{Cooperative Multi-Agent Reinforcement Learning with Large Language Models}

The development of large models, especially large language models~\cite{zhao2023survey}, has gained widespread attention and application in various fields in recent years. Some recent works are exploring universal decision-making large models~\cite{ye2023foundation, yang2023foundation} and applying them in different contexts. In SARL tasks, research works like GATO~\cite{reed2022generalist}, DreamerV3~\cite{hafner2023mastering}, and DT~\cite{chen2021decision} have achieved surprising results in many task scenarios. These works leverage the powerful expressive capabilities of existing technologies such as Transformer~\cite{vaswani2017attention, hu2022transforming}.
On the other hand, some recent works are attempting to learn universal decision-making large models for MASs. For example, MADT~\cite{meng2021offline} promotes research by providing a large-scale dataset and explores the application of DT in MARL environments. MAT~\cite{wen2022multiagent} studies an effective large model method to transform MARL into a single-agent problem, aiming to map the observation sequences of agents to optimal action sequences, demonstrating superior performance in multiple task scenarios compared to traditional methods. Addressing the information of entities in multi-agent environments, literature~\cite{wang2023multiagent} proposed MAGENTA, a study orthogonal to previous time-series modeling. MADiff~\cite{zhu2023madiff} and DOM2~\cite{li2023beyond} introduced generative diffusion models to MARL, promoting collaboration in various scenarios. SCT~\cite{li2023selfconfirming} accelerates multi-agent online adaptation using a Transformer model. Literature~\cite{park2023generative} constructed a humanoid multi-agent environment, "West World," to simulate and test MASs in large-scale scenarios.

Additionally, with the development of large language models represented by ChatGPT, some works are attempting to promote multi-agent collaboration through language models. For instance, EnDi~\cite{ding2023entity} uses natural language to enhance the generalization ability of MASs. InstructRL~\cite{DBLP:conf/icml/HuS23} allows humans to obtain the desired agent strategy through natural language instructions. SAMA~\cite{li2023semantically} proposes semantically aligned multi-agent collaboration, automatically assigning goals to MASs using pre-trained language prompts, achieving impressive collaboration results in various scenarios. HAPLAN~\cite{guan2023efficient} utilizes large language model like ChatGPT to bride the gap between human and AI for efficient coordination.
ProAgent~\cite{zhang2023proagent} introduces an efficient human-machine collaboration framework that leverages large language models for teammate behavior prediction, achieving the best collaborative performance in human-machine collaboration tasks. On the other hand, some works apply multi-agent collaboration methods to enhance the capabilities of large language models~\cite{liang2023encouraging, chan2023chateval, wu2023autogen}.
However, due to reasons such as complex interactions, the field of universal decision-making large models for multi-agents is currently less explored. Challenges such as learning a universal multi-agent decision-making large model that can generalize with zero or few shot across various scenarios or quickly adapt to new tasks through fine-tuning are worth researching.

\section{Summary and Prospect}
This paper focuses on the development and research of cooperative MARL, progressing from classical closed environments to open environments that align with real-world applications. It provides a comprehensive introduction to reinforcement learning, multi-agent systems, multi-agent reinforcement learning, and cooperative multi-agent reinforcement learning. Summarizing different research directions, it extracts and condenses the key research focuses on MARL in classical environments.
Despite the success of many closed-environment MARL algorithms based on the closed-world assumption, their application in the real world remains limited. This limitation is largely attributed to the lack of targeted research on the characteristics of open environments. There exists a substantial gap between the current methods and the actual empowerment of daily life. To overcome the challenges posed by the complexity, dynamism, and numerous constraints in open environments, future research in cooperative MARL could address the following aspects. This will trigger more attention and exploration of MARL in open environments, enabling better application of cooperative MASs in real-world settings to enhance human life.

\begin{itemize}
    \item \textbf{Solutions of Multi-Agent Coordination Issues in Classical Closed Environments}: Algorithms for Cooperative Multi-agent Reinforcement Learning in classical environments serve as the foundation for transitioning to open environments. Improving the coordination performance in closed environments enhances the broader applicability potential of these systems. However, challenges persist in large-scale scenarios with numerous intelligent agents, such as efficient policy optimization~\cite{xiong2023sampleefficient}, and balancing the relationship between distributed and centralized aspects during training and execution. These issues require careful consideration and resolution in future research.
    
    \item \textbf{Theoretical Analysis and Framework Construction in Open Environments}: Open environments present more stringent conditions and greater challenges compared to closed ones. While some efforts have utilized heuristic rules to design the openness of machine learning environments~\cite{zhou2022open}, the establishment of a comprehensive framework, including a well-defined concept of openness in multi-agent coordination, definitions of environmental openness, and the performance boundaries of algorithms, is a crucial area for future research.
    
    \item \textbf{Construction of Testing Environments for Cooperative MASs in Open Environments}: Despite ongoing research on robustness in open environments for cooperative MARL, benchmark testing often involves modifications in classical closed environments. These approaches lack compatibility with various challenges posed by different open-ended scenarios. Future research could focus on constructing testing environments that evaluate eleven identified aspects, providing a significant boost to the study of cooperative MASs in open environments.
    
    \item \textbf{Development of General Decision-Making Large Language Models for Multi-agent Systems in Open Environments}: Large models, especially large language models~\cite{zhao2023survey}, have garnered attention and applications in various fields. Some works explore decision-making large language models in multi-agent settings~\cite{yang2023foundation,zhang2023rladapter}, yet there is still a considerable gap in research. Future investigations could focus on learning universal decision-making large language models for MASs that generalize across diverse task scenarios, achieving zero or few-shot generalization, or rapid adaptation to unknown task domains through fine-tuning.
    
   \item \textbf{Application and Implementation of Cooperative Multi-agent Reinforcement Learning in Real-world Scenarios}: While the efficient cooperative performance of MARL in classical environments holds great application potential, most studies are limited to testing in simulators or specific task scenarios~\cite{li2019reinforcement}. There is still a considerable distance from real-world social applications and current needs. The primary goal of research in cooperative MARL in open environments remains the application of algorithms to human life and the promotion of societal progress. In the future, exploring how to safely and efficiently apply MARL algorithms in areas such as large-scale autonomous driving, smart cities, and massive computational resource scheduling is a topic worthy of discussion.
\end{itemize}

\section*{Acknowledgments}
 We would like to thank Rongjun Qin, Fuxiang Zhang, Chenghe Wang, Yichen Li, Ke Xue, Chengxing Jia, Feng Chen, and Zhichao Wu for their helpful discussions and support. 

\bibliographystyle{unsrt}
\bibliography{ref}
\end{document}